\documentclass[journal,twoside]{IEEEtran}

\pdfoutput=1
\usepackage{graphicx}
\usepackage[utf8]{inputenc}

\usepackage{amsmath}
\usepackage{wasysym}
\usepackage{ifthen}
\usepackage{xspace}
\usepackage{xcolor}
\usepackage{booktabs}	
\usepackage{commath}

\newcommand{\cnn}{\emph{ \textcolor{red}{[citation needed]}} }
\newcommand{\cn}[1]{\textcolor{red}{[\emph{#1}]}}
\newcommand\todo[1]{\textcolor{red}{#1}}
\newcommand*{\eg}{e.g.\@\xspace}
\newcommand*{\ie}{i.e.\@\xspace}
\newcommand*{\cf}{cf.\@\xspace}
\newcommand\refeq[1]{eq.~\ref{#1}}
\newcommand\refeqs[1]{eqs.~\ref{#1}}
\newcommand\refig[1]{fig.~\ref{#1}}
\newcommand\reFig[1]{Fig.~\ref{#1}}
\newcommand\refigs[1]{figs.~\ref{#1}}
\newcommand\reFigs[1]{Figs.~\ref{#1}}
\newcommand{\gr}{$\gamma$-ray\xspace}
\newcommand{\grs}{$\gamma$-rays\xspace}
\newcommand{\tup}{\textsuperscript}
\newcommand{\tlow}{\textsubscript}

\usepackage[nolist]{acronym}

\usepackage{cite}

\providecommand\rapido{false}
\providecommand\publicar{false}
\ifthenelse{ \equal{\publicar}{false} }
{\usepackage{hyperref}}
{\usepackage[bookmarks=false]{hyperref}}
\usepackage{tikz}
\usetikzlibrary{patterns,plotmarks}
\usetikzlibrary{decorations.pathmorphing,arrows,shapes,positioning,patterns,calc,through,backgrounds}
\usetikzlibrary{calc}

\newcommand{\tikzAngleOfLine}{\tikz@AngleOfLine}
  \def\tikz@AngleOfLine(#1)(#2)#3{
  \pgfmathanglebetweenpoints{
    \pgfpointanchor{#1}{center}}{
    \pgfpointanchor{#2}{center}}
  \pgfmathsetmacro{#3}{\pgfmathresult}
  }
  
\usepackage{tikz-3dplot} 

\definecolor{paperblue} {HTML}{9fd7ff}
\definecolor{papergreen}{HTML}{b5e04b}
\definecolor{papergray} {HTML}{c3c3c3}
\definecolor{paperwalls}{HTML}{bfbfbf}
\definecolor{paperbrown}{HTML}{bf8040}
\definecolor{paperbones}{HTML}{ffcc33}

\usepackage{tikz-dimline,pgf}
\pgfplotsset{compat=1.14}
\makeatletter
\newcommand{\Distance}[3]{
\tikz@scan@one@point\pgfutil@firstofone($#1-#2$)\relax  
\pgfmathsetmacro{#3}{int(round(veclen(\the\pgf@x,\the\pgf@y)/1000/0.0283465*0.99626))}
}
\makeatother
 \usepackage{onimage}

\usepackage{nicefrac}

\DeclareMathOperator\erf{erf}
\newcommand{\me}{\mathrm{e}}
\DeclareMathOperator*{\argmax}{argmax}
\DeclareMathOperator*{\argmin}{argmin}

\usepackage{tabularx}
\usepackage{colortbl}
\colorlet{helpful}{lime!70}
\colorlet{harmful}{red!30}
\colorlet{internal}{yellow!20}
\colorlet{external}{cyan!30}
\colorlet{S}{helpful!50!internal}
\colorlet{W}{harmful!50!internal}
\colorlet{O}{helpful!50!external}
\colorlet{T}{harmful!50!external}
\colorlet{opport}{lime!70!cyan!90}
\colorlet{threat}{gray!70}

\newcommand{\textcn}{\rotatebox[origin=c]{90}{\parbox[t]{3cm}{\centering Internal origin\\ \tiny (product\slash company attributes)\par}}}
\newcommand{\textdn}{\rotatebox[origin=c]{90}{\parbox[b]{3cm}{\centering External origin\\ \tiny (environment\slash market attributes)\par}}}

\newcommand{\back}[1]{}
\newcommand\mycolor[1]{\cellcolor{#1}}
\newcolumntype{C}[1]{>{\centering\arraybackslash}m{#1}}

\makeatletter
\newcommand\notsotiny{\@setfontsize\notsotiny\@vipt\@viipt}
\makeatother
\usepackage{eso-pic}
\AddToShipoutPictureBG{
  \AtPageLowerLeft{
    \raisebox{1.25\baselineskip}{\makebox[\paperwidth]{\begin{minipage}{8.5in}\centering
\notsotiny
      2469--7311\copyright~2019 IEEE. Personal use is permitted, but republication/redistribution requires IEEE permission. See \url{http://www.ieee.org/publications_standards/publications/rights/index.html} for more information. 
    \end{minipage}}}
  }
  \AtPageUpperLeft{
    \raisebox{-1.25\baselineskip}{\makebox[\paperwidth]{\begin{minipage}{8.5in}\centering
\notsotiny
      This article has been accepted for publication in a future issue of this journal, but has not been fully edited. Content may change prior to final publication. Citation information: \href{https://doi.org/10.1109/TRPMS.2019.2930362}{\textsc{DOI 10.1109/TRPMS.2019.2930362}},\\IEEE Transactions on Radiation and Plasma Medical Sciences 
    \end{minipage}}}
  }
} 
\begin{acronym}[KVI-CART]
\acro{1D}{One-Dimensional}
\acro{2D}{Two-Dimensional}
\acro{3D}{Three-Dimensional}
\acro{ADC}{Analog to Digital Converter}
\acro{AGOR}{Accélérateur Groningen ORsay}
\acro{BbCc}{BGO block Compton camera}
\acro{BGO}{Bismuth Germanium Oxide - Bi\tlow{4}Ge\tlow{3}O\tlow{12}}
\acro{BP}{Back Projection}
\acro{C230}{Cyclone\textsuperscript{\textregistered}\,230}
\acro{CERN}{Conseil Européen pour la Recherche Nucléaire}
\acro{CFD}{Constant Fraction Discriminator}
\acro{CPGM}{Coaxial Prompt Gamma-ray Monitoring}
\acro{CT}{Computed Tomography}
\acro{CTR}{Coincidence Time Resolution}
\acro{CSDA}{Continuous Slowing Down Approximation}
\acro{CZT}{Cadmium zinc telluride - CdZnTe}
\acro{DCEU}{Digital Counter Electronic Unit}
\acro{DDR3}{Double Data Rate type 3}
\acro{ELBE}{Electron Linear accelerator for beams with high Brilliance and low Emittance}
\acro{EPT}{Electron-Positron Tracking}
\acro{FPGA}{Field Programmable Gate Array}
\acro{FWHM}{Full Width at Half Maximum}
\acro{GEVI}{Gamma Electron Vertex Imaging}
\acro{GPIB}{General Purpose Interface Bus}
\acro{GSO}{Cerium-doped gadolinium oxyorthosilicate - Gd\tlow{2}SiO\tlow{5}:Ce}
\acro{HV}{High Voltage}
\acro{KVI-CART}{Kernfysisch Versneller Instituut - Center for Advanced Radiation Technology}
\acro{LLUMC}{Loma Linda University Medical Center}
\acro{LSO}{Cerium-doped lutetium oxyorthosilicate - Lu\tlow{2}SiO\tlow{5}:Ce}
\acro{MAW}{Moving Average Window}
\acro{MC}{Monte Carlo}
\acro{MGH}{Massachusetts General Hospital}
\acro{MLEM}{Maximum-Likelihood Expectation-Maximisation}
\acro{MOSFET}{Metal-Oxide Semiconductor Field-Effect Transistor}
\acro{MRI}{Magnetic Resonance Imaging}
\acro{NIM}{Nuclear Instrumentation Module}
\acro{NPDF}{Normalised Probability Density Function}
\acro{OAR}{Organ At Risk}
\acro{PA}{Preamplifier}
\acro{PBS}{Pencil Beam Scanning}
\acro{PCIe}{Peripheral Component Interconnect Express}
\acro{PET}{Positron Emission Tomography}
\acro{PGI}{Prompt Gamma-ray Imaging}
\acro{PGPI}{Prompt Gamma-ray Peak Integration}
\acro{PGS}{Prompt Gamma-ray Spectroscopy}
\acro{PGT}{Prompt Gamma-ray Timing}
\acro{PMMA}{Polymethyl Methacrylate - [C\tlow{5}O\tlow{2}H\tlow{8}]\tlow{n}}
\acro{PMT}{Photomultiplier Tube}
\acro{QDC}{Charge to Digital Converter}
\acro{RBNS}{Recursive Bisection Neutron Subtraction}
\acro{RF}{Radio Frequency}
\acro{SDEP}{Single and Double Escape Peaks}
\acro{simBox}{simple Box}
\acro{SNIP}{Sensitive Nonlinear Iterative clipping Peak}
\acro{SOBP}{Spread Out Bragg Peak}
\acro{SWOT}{Strengths, Weaknesses, Opportunities and Threats}
\acro{TDC}{Time to Digital Converter}
\acro{UFH}{University of Florida Health Proton Therapy Institute}
\acro{USA}{United States of America}
\acro{UPTD}{Universit\"ats Protonen Therapie Dresden}
\acro{VME}{VERSAmodule Eurocard}
\acro{WET}{Water Equivalent Thickness}
\end{acronym}

\newcommand{\ADC}{\ac{ADC}\@\xspace}
\newcommand{\AGOR}{\ac{AGOR}\@\xspace}
\newcommand{\BbCc}{\ac{BbCc}\@\xspace}
\newcommand{\BGO}{\ac{BGO}\@\xspace}
\newcommand{\BP}{\ac{BP}\@\xspace}
\newcommand{\CERN}{\ac{CERN}\@\xspace}
\newcommand{\CFD}{\ac{CFD}\@\xspace}
\newcommand{\CPGM}{\ac{CPGM}\@\xspace}
\newcommand{\CT}{\ac{CT}\@\xspace}
\newcommand{\CTR}{\ac{CTR}\@\xspace}
\newcommand{\CSDA}{\ac{CSDA}\@\xspace}
\newcommand{\CZT}{\ac{CZT}\@\xspace}
\newcommand{\ELBE}{\ac{ELBE}\@\xspace}
\newcommand{\EPT}{\ac{EPT}\@\xspace}
\newcommand{\FPGA}{\ac{FPGA}\@\xspace}
\newcommand{\FWHM}{\ac{FWHM}\@\xspace}
\newcommand{\GEVI}{\ac{GEVI}\@\xspace}
\newcommand{\GSO}{\ac{GSO}\@\xspace}
\newcommand{\HZDR}{\ac{HZDR}\@\xspace}
\newcommand{\HV}{\ac{HV}\@\xspace}
\newcommand{\KVI}{\ac{KVI-CART}\@\xspace}
\newcommand{\LSO}{\ac{LSO}\@\xspace}
\newcommand{\MAW}{\ac{MAW}\@\xspace}
\newcommand{\MC}{\ac{MC}\@\xspace}
\newcommand{\MGH}{\ac{MGH}\@\xspace}
\newcommand{\MLEM}{\ac{MLEM}\@\xspace}
\newcommand{\NPDF}{\ac{NPDF}\@\xspace}
\newcommand{\OAR}{\ac{OAR}\@\xspace}
\newcommand{\PA}{\ac{PA}\@\xspace}
\newcommand{\PBS}{\ac{PBS}\@\xspace}
\newcommand{\PET}{\ac{PET}\@\xspace}
\newcommand{\PGI}{\ac{PGI}\@\xspace}
\newcommand{\PGPI}{\ac{PGPI}\@\xspace}
\newcommand{\PGS}{\ac{PGS}\@\xspace}
\newcommand{\PGT}{\ac{PGT}\@\xspace}
\newcommand{\PMMA}{\ac{PMMA}\@\xspace}
\newcommand{\PMT}{\ac{PMT}\@\xspace}
\newcommand{\PMTs}{\acp{PMT}\@\xspace}
\newcommand{\QDC}{\ac{QDC}\@\xspace}
\newcommand{\RBNS}{\ac{RBNS}\@\xspace}
\newcommand{\RF}{\ac{RF}\@\xspace}
\newcommand{\SDEP}{\ac{SDEP}\@\xspace}
\newcommand{\simBox}{\ac{simBox}\@\xspace}
\newcommand{\SNIP}{\ac{SNIP}\@\xspace}
\newcommand{\SWOT}{\ac{SWOT}\@\xspace}
\newcommand{\TDC}{\ac{TDC}\@\xspace}
\newcommand{\UPTD}{\ac{UPTD}\@\xspace}
\newcommand{\USA}{\ac{USA}\@\xspace}
\newcommand{\VME}{\ac{VME}\@\xspace}
\newcommand{\WPE}{\ac{WPE}\@\xspace} 
\begin{document}
\bstctlcite{IEEEexample:BSTcontrol}

\title{Compact Method for Proton Range Verification\\Based on Coaxial Prompt Gamma-Ray Monitoring:\\a Theoretical Study}

\author
{
F.~Hueso-González
and T.~Bortfeld
\thanks
{
F.~Hueso-González and T.~Bortfeld are with Department of Radiation Oncology, Massachusetts General Hospital and Harvard Medical School, Boston, MA 02114, United States of America.
}
\thanks
{
Contact: fhuesogonzalez@mgh.harvard.edu. This work was supported in part by the Federal Share of program income earned by Massachusetts General Hospital on C06-CA059267, Proton Therapy Research and Treatment Center.
}
\thanks{Manuscript submitted February 28, 2019; accepted July 18, 2019.}
}

\markboth{Transactions on Radiation and Plasma Medical Sciences,~Vol.~4, No.~2, March~2020}
{Hueso-González \MakeLowercase{\textit{et al.}}: Proton Range Verification Based on Coaxial Prompt Gamma-Ray Monitoring}

\IEEEspecialpapernotice{Special Issue on Particle Therapy}

\maketitle

\begin{abstract}
Range uncertainties in proton therapy hamper treatment precision. Prompt gamma-rays were suggested 16 years ago for real-time range verification, and have already shown promising results in clinical studies with collimated cameras. Simultaneously, alternative imaging concepts without collimation are investigated to reduce the footprint and price of current prototypes. In this manuscript, a compact range verification method is presented. It monitors prompt gamma-rays with a single scintillation detector positioned coaxially to the beam and behind the patient. Thanks to the solid angle effect, proton range deviations can be derived from changes in the number of gamma-rays detected per proton, provided that the number of incident protons is well known. A theoretical background is formulated and the requirements for a future proof-of-principle experiment are identified. The potential benefits and disadvantages of the method are discussed, and the prospects and potential obstacles for its use during patient treatments are assessed. The final milestone is to monitor proton range differences in clinical cases with a statistical precision of 1 mm, a material cost of 25000 USD and a weight below 10 kg. This technique could facilitate the widespread application of in vivo range verification in proton therapy and eventually the improvement of treatment quality.
\end{abstract}

\begin{IEEEkeywords}
proton therapy, range verification, prompt gamma rays, radiation detectors, coaxial, compact.
\end{IEEEkeywords}

\IEEEpeerreviewmaketitle 
\section{Introduction}
\IEEEPARstart{T}{he} number of cancer patients treated worldwide with proton beams has increased from $\sim$6000 per year in 2007 to $\sim$20000 per year in 2017 \cite{ptcog_17}. Paired with the advance in beam delivery and treatment planning techniques, the number of hospitals with a proton therapy facility increases annually, up to $\sim$80 in 2019. The essential advantage of proton beams over conventional radiotherapy is their increased ionization density (Bragg peak) and sharp distal falloff near their maximum penetration depth \cite{review_pt,tian_review_pt}. It is estimated that $\sim$10\,\% of all cancer patients, particularly children, could benefit from proton therapy \cite{amaldi_review,glimelius_number}. 
However, there is no standardized or commercial method applied in all proton therapy facilities to verify in real-time where protons stop within the patient \cite{range_review}. 
This inherent range uncertainty \cite{andreo_range} limits to some extent the potential of protons to conform the dose to the tumor. Currently, it forces the application of field patching techniques and conservative safety margins \cite{albertini_margins} during treatment planning, of up to $\sim$10\,mm \cite{hueso_fio_cc_pgt}. A robust treatment plan \cite{unkelbach_robust} can ensure tumor coverage even in the case of proton range deviation, but at the cost of a higher integral dose to normal surrounding tissue \cite[Fig.~5]{polf_margins}.

Thanks to the efforts of many research institutions during the last decades, several solutions towards in vivo range verification have been proposed and tested \cite{range_review,parodi_range_review}. Two examples thereof are \PET \cite{parodi_pet} and \PGI \cite{smeets_slit}. The first one has been extensively tested in clinical settings, but is challenged by the correlation of activity to dose as well as the metabolic washout effect \cite{kraan_review, parodi_range_review}, except in the case of in-beam \PET of short-lived nuclides \cite{buitenhuis_ibpet}.

The second technique, proposed in 2003 \cite{jongen_pg}, has shown promising advances in recent years \cite{krimmer_review}. First clinical tests were done with a knife-edge slit \PGI camera \cite{richter_slit,xie_slit}, and a clinical study with a \PGS camera \cite{hueso_pgs} is foreseen. Both camera prototypes rely on a passive collimation of prompt \grs (with energies up to $\sim$6\,MeV) and are able to detect the proton range in clinical conditions with a few millimeters precision.

The higher costs of proton therapy with respect to conventional radiotherapy also limit its applicability \cite{tian_review_pt}. There have been efforts from research institutions and industrial partners to reduce the size of proton accelerators \cite{schippers_size,farr_horizons} as well as the overall price of facilities \cite{slopsema_size,bortfeld_affordable}. Likewise, there is a trend in the field of proton range verification to reduce both the cost and size of currently available prototypes by proposing innovative detection approaches. Collimated systems like the \PGI and \PGS cameras deploy thick tungsten collimators, that contribute substantially to the overall system weight and footprint. Indirectly, it also increases the costs due to the need of an accurate positioning and rotation system sustaining this weight near the patient. In response to this drawback, uncollimated range verification systems have been proposed and are under investigation \cite{parodi_iono,golnik_pgt,krimmer_pgpi,draeger_polaris}. Their smaller footprint and weight could eventually facilitate the clinical translation.

This article is framed in this context: The need for providing a range verification system in every treatment room of every proton therapy facility at an affordable cost. To that extent, the experience from the current clinical prototypes is essential, but innovative approaches have to be developed to reduce their weight, price and complexity, while still addressing the required precision and associated technical challenges \cite{pausch_needs}.

The dependency between the number of prompt \grs detected per proton (with uncollimated systems) and the beam range was observed in \cite[Fig.~9]{golnik_pgt} within the context of \PGT, in \cite[Fig.~10]{hueso_bbcc} using a Compton camera setup, as well as in \MC simulations \cite[Fig.~7]{eunsin_angular}. This information has been incorporated into the proton range reconstruction algorithm in \cite[subsection~3.1.4, Fig.~3.14]{hueso_phd}. This methodology was later isolated and named as \PGPI \cite{krimmer_pgpi}.

The aim of this paper is to propose a compact range verification method -- {\CPGM} -- that exploits the aforementioned phenomenon. The innovation is the positioning of the detector coaxial to the beam axis and behind the treated area. This enables the determination of proton range deviations with a single detector solely based on the number of \grs measured and the total incident protons. The covered solid angle will be maximum directly in front of the detector and decreases with the inverse of the square distance (at first order). Consequently, a proton range overshoot caused by \eg an unexpected air cavity leads to an increased number of \grs measured per incoming proton, and vice versa for a range undershoot.

This manuscript is organized as follows. Section~\ref{sec:theory} describes a simple mathematical model and solid angle equations of coaxial radiation detection, that serves to illustrate the proposed method. In section~\ref{sec:results}, the analytical model is evaluated for a clinically relevant test case, that could be verified in a future proof-of-principle experiment. The implications, benefits and disadvantages of \CPGM in terms of clinical translation are discussed in section~\ref{sec:discussion}, and the main conclusions are summarized in section~\ref{sec:conclusions}. 
\section{Theory}\label{sec:theory}

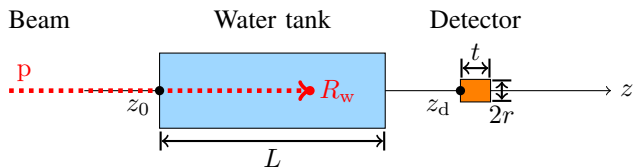
\begin{figure}[h]
\centering
\begin{tikzpicture}

	\coordinate (A) at (1,-0.67);
	\coordinate (B) at (4,-0.67);

    \draw [->] (0,0) -- (7,0) node (xaxis) [right] {$z$};
    \draw (5.2,0) node [above=0.7] {Detector};

	\draw [fill=paperblue]   (1,-0.5) rectangle (4,0.5);    
	\draw [fill=orange] (5,-0.15) rectangle (5.4,0.15);    
	
	\draw (2.5,0) node [above=0.7] {Water tank};

	\draw[line width=1.75pt,dotted,red,->] (-1,0) -- (3,0);    
	\draw[red] (-0.8,0) node [above] {$\mathrm{p}$}; 
    \draw (-0.6,0) node [above=0.7] {Beam};
    
    \draw[thick,|<->|] (A) -- (B);
    \fill[black] (2.5,-0.65) node [anchor=north] {$L$};
    \fill[black] (1,0) circle (1.75pt) node [anchor=north east] {$z_0$};
    \fill[red]   (3,0) circle (1.75pt) node [anchor=west] {$R_\mathrm{w}$};
    \fill[black] (5,0) circle (1.75pt) node [anchor=north east] {$z_\mathrm{d}$};
    
    \draw[thick,|<->|] (5,0.3) -- (5.4,0.3);
    \fill[black] (5.2,0.28) node [anchor=south] {$t$};
    
        \draw[thick,|<->|] (5.55,0.15) -- (5.55,-0.15);
    \fill[black] (5.55,0) node [below=0.5ex] {$2r$};
    
\end{tikzpicture}
 \caption{Schematics of the detector setup coaxial to the proton beam incidence direction $z$ and behind a water target of length $L$. The phantom entrance is at position $z_0$, whereas the detector front face is at position $z_\mathrm{d}> z_0+L$. The detector is a monolithic cylindrical crystal of radius $r$ and thickness $t$. The proton range within the water tank is $R_\mathrm{w}<L$.}
\label{fig:setup}
\end{figure}

To provide a theoretical framework showing the potential of the \CPGM technique, we investigate a simplified setup with a homogeneous water phantom and a cylindrical monolithic detector located downstream and aligned with the proton beam axis, \cf \refig{fig:setup}.

In the following subsections, we formulate an approximate mathematical model of the \gr emission, transmission, solid angle coverage and \gr detection, as well as an estimate of the radiation background, detector count rates and sensitivity to range errors in various clinical scenarios. An overview of the variables and parameters defined throughout this section is presented in Table~\ref{tab:params}.

\subsection{Emission}

We assume an analytical \ac{1D} model of the prompt \gr emission (including all energies) characterized by a flat profile and a sharp distal falloff near the proton range $R_\mathrm{w}$. This is a coarse approximation of the energy-integrated depth-emission profile \cite[Fig.~6]{biegun_tof}, \cite[Fig.~7]{verburg_pg}. The expression of \grs emitted per unit length and incident proton yields:
\begin{equation}
\od{N_\gamma}{z}(z;R_\mathrm{w})\approx\bar{g}_\lambda\cdot\frac{1}{2}\left(1+\erf[\bar{f}(R_\mathrm{w}-(z-z_0))]\right)
\label{eq:dNe}
\end{equation}
for $z\in(z_0,z_0+L)$, \cf \refig{fig:setup}. The integral number of emitted \grs can be approximated for $\bar{f}^{-1}\ll R_\mathrm{w}$ by:
\begin{equation}
N_\gamma(R_\mathrm{w})=\int\limits_{z_0}^{z_0+L} \od{N_\gamma}{z}(z)\dif z \approx \bar{g}_\lambda R_\mathrm{w}
\label{eq:gammaint}
\end{equation}

The (approximate) value $\bar{g}_\lambda\approx6\times10^{-4}\,\text{mm}^{-1}$ is the energy-integrated prompt \gr yield, namely the number of \grs emitted per proton traversing a unit length in water.
$\bar{f}\approx0.3\,\text{mm}^{-1}$ is an empirical constant to reproduce the shape of the distal drop-off of the \gr emission at the proton range $R_\mathrm{w}$. For simplicity, it is considered to be independent from the initial proton energy. For $\bar{f}\to\infty$, the error function becomes a step function and the emission profile is a simple rectangular function \cite[Fig.~3]{golnik_pgt}.

\subsection{Transmission}
The prompt \grs emitted towards the detector might be absorbed, scattered or undergo pair production within the water phantom. The transmission probability is approximated as a function of the energy-dependent absorption coefficient $\mu_\mathrm{w}(E)$ \cite{nist_xcom} and the traversed path. As a first order approximation, we use a value $\bar{\mu}_\mathrm{w}\approx10^{-4}\,\text{mm}^{-1}$ that is the sum of the photoelectric absorption and pair production coefficients averaged over the energy range of interest. For the setup described in \refig{fig:setup}, the transmission probability $\tau(z)$ of a ray emitted at a depth $z\in(z_0,z_0+L)$ and traveling towards the detector is given by:
\begin{equation}
\tau(z)=\exp\left[-\bar{\mu}_\mathrm{w}\left(L-(z-z_0)\right)\right]
\label{eq:trans}
\end{equation}

Compton scattering is not included, as high energy \grs are likely to scatter in forward direction \cite[Fig.~2.4]{hueso_phd} and thus might still impinge the detector. On the other hand, annihilation photons (following pair production) are less likely to escape the phantom and reach the detector, due to the three times higher attenuation coefficient in water compared to 4\,MeV photons \cite{nist_xcom}, as well as due to their isotropic emission distribution; hence their contribution is neglected. 

\subsection{Solid angle}
The influence of detector position and solid angle covered with respect to an irradiation source has been comprehensively studied in \cite{solid_angle,conway_solid}, and was investigated specifically for the \PGI field by means of \MC simulations \cite{zarifi_angle,eunsin_angular,polf_distance}. Here, we adopt the analytical solution for an ideal cylindrical detector of thickness $t$ and radius $r$, aligned with the beam axis. For a \gr emitted isotropically at a depth $z\in(z_0,z_0+L)$, the relative solid angle $\Omega/4\pi$ subtended by the front face of the detector located at position $z_\mathrm{d}$ yields:
\begin{equation}
\frac{\Omega}{4\pi}(z;z_\mathrm{d})=\frac{1}{2}\left(1-\sqrt{\frac{1}{1+r^2/(z_\mathrm{d}-z)^2}}\right)
\label{eq:solid}
\end{equation}

\subsection{Detection}
For a \gr crossing the detector, the interaction probability depends on the crystal thickness $t$ and the energy-dependent attenuation coefficient $\mu_\mathrm{d}(E)$ \cite{nist_xcom}. In the paraxial approximation, and using an averaged attenuation coefficient $\bar{\mu}_\mathrm{d}$, the interaction probability (including scattering) yields:
\begin{equation}
\eta=1-\exp[-\bar{\mu}_\mathrm{d}t]
\label{eq:detection}
\end{equation}

A fast scintillation crystal like LaBr\tlow{3} with an average attenuation coefficient of $\bar{\mu}_\mathrm{d}\approx2\times10^{-2}\,\text{mm}^{-1}$ \cite{nist_xcom} and a decay time $t_\mathrm{d}$ of $16$\,ns \cite{glodo_prbr3} is chosen as detector. The monolithic scintillation crystal has a diameter ($2r$) of $1$\,'' and a thickness ($t$) of $1\nicefrac{1}{2}$\,''.

The number of prompt \grs $\hat{N}_\mathrm{d}$ measured by the detector at a position $z_\mathrm{d}$ for each incident proton with a range $R_\mathrm{w}$ follows from the multiplication of production $\dif{N_\gamma}/\dif{z}$, transmission $\tau$, solid angle $\Omega$ and detection $\eta$ probabilities, summed across the target:
\begin{equation}
\hat{N}_\mathrm{d}(R_\mathrm{w},z_\mathrm{d})\approx\int\limits_{z_0}^{z_0+L} \od{N_\gamma}{z}(z;R_\mathrm{w})\cdot\tau(z)\cdot\frac{\Omega}{4\pi}(z;z_\mathrm{d})\cdot\eta\cdot\dif z
\label{eq:gamma_counts}
\end{equation}

For a pencil beam delivering $N_\mathrm{p}$ protons, the total prompt \gr counts $N_\mathrm{d}$ are given by:
\begin{equation}
N_\mathrm{d}=\hat{N_\mathrm{d}} \cdot N_\mathrm{p}
\label{eq:spot_counts}
\end{equation}

It is common practice in \PBS to aggregate data from neighboring spots \cite{xie_slit,nenoff_slit,hueso_pgs} in order to achieve around $N_\mathrm{p}\sim10^8$ protons per cluster. Since the lateral spread of the pencil beam close to the proton range is often larger than the lateral separation between pencil beams, an improvement in statistical precision can be achieved without significant loss in lateral resolution.

In the case of accurately characterized phantoms, and well-known prompt \gr yield, detector efficiency and position, the absolute proton range $R_\mathrm{w}$ can be reconstructed by minimizing the deviation between measured $M_\mathrm{d}$ and theoretically expected (\refeq{eq:gamma_counts}) prompt \gr counts:
\begin{equation}
\underset{R_\mathrm{w}}{\argmin}\left| M_\mathrm{d}-\hat{N_\mathrm{d}}(R_\mathrm{w}) \cdot N_\mathrm{p} \right|
\label{eq:range}
\end{equation}

\subsection{Radiation background}\label{ss:background}
In addition to the prompt \gr signal, a large radiation background is impinging on the detector. It is induced by neutron interactions as well as material activation (\eg positron emitting isotopes). Due to the time correlation between the \gr signal and the cyclotron \RF, the background can be reduced \cite{peterson_g4,biegun_tof,cambraia_background}, or subtracted via dedicated algorithms using timing and energy filters \cite{hueso_pgs}. However, even then, the background events are still important for quantifying the overall detector load. The actual background is difficult to model and depends on the incident proton energy, the target composition, as well as the detector location and other elements in the treatment room. Based on previous measurements in a clinical scenario \cite{pinto_absolute,hueso_pgt,hueso_pgs} as well as simulations \cite[Fig.~6]{biegun_tof}, the average ratio between overall events and prompt \grs can be roughly estimated with $k_\gamma\approx 2.5$ for the current setup in coaxial orientation and without collimator, \cf \refig{fig:setup}. It is assumed that an event can only be detected if it deposits at least 50\,keV in the detector.

\subsection{Detector count rate}
The beam current in a proton therapy facility is a critical factor challenging the electronics and affecting the feasibility of \PGI devices \cite{hueso_bbcc,pausch_needs}. For the widespread Cyclone\textsuperscript{\textregistered}\,230 (C230) isochronous cyclotron of IBA (Louvain-la-Neuve, Belgium), the instantaneous proton beam current $I_\mathrm{p}$ is approximately constant at 2\,nA \cite{slit_clinical_curr} at the beam exit window. Correspondingly, the overall detector count rate $\dot{N}_\mathrm{d,all}$ including background is:
\begin{equation}
\dot{N}_\mathrm{d,all}\approx k_\gamma\cdot\hat{N_\mathrm{d}}\cdot \frac{I_\mathrm{p}}{e}
\label{eq:dload}
\end{equation}
where $e$ is the elementary charge.

\subsection{Test case}
We defined the irradiation of a brain tumor from a horizontal incidence angle and with the patient couch perpendicular to the beam axis as benchmark case representative for a clinical proton treatment. This treatment site is included among the accepted indications \cite{testa_phd,astro_mp} of proton therapy, and the chosen beam incidence and couch angles are frequent in clinical treatment plans \cite{susu_gantry,taasti_margins}.

To resemble this geometry in our model, \cf \refig{fig:setup}, we choose the water tank length $L$ to be equal to the average human head breadth of 150\,mm \cite{zhuang_head_width}. The detector is located at a distance $z_\mathrm{d}=$\,152\,mm from the beam entrance point $z_0=$\,0\,mm. 

\subsection{Sensitivity to range errors}

The sensitivity of the \CPGM method to proton range deviations can be studied analytically by calculating the derivatives of \refeq{eq:gamma_counts} with respect to parameters of interest. For example, the relative variation $\Delta_{N}$ of the number of detected \grs $\hat{N}_\mathrm{d}$ for an error $\delta_R$ in the proton range in water $R_\mathrm{w}$, caused by an error in the initial proton energy, is given by:
\begin{eqnarray}
\Delta_{N}& \equiv &\frac{1}{\hat{N}_\mathrm{d}}\frac{\partial\hat{N}_\mathrm{d}}{\partial R_\mathrm{w}}(R_\mathrm{w})\cdot\delta_R \label{eq:gamma_counts_dif}\\
& = &\frac{\delta_R}{\hat{N}_\mathrm{d}}\int\limits_{z_0}^{z_0+L} \frac{\bar{g}_\lambda \bar{f}}{\sqrt{\pi}} \me^{-\bar{f}^2 R_\mathrm{w}^2(z-z_0)^2}\tau\cdot\frac{\Omega}{4\pi}\cdot\eta\cdot\dif z \label{eq:deltaN}
\end{eqnarray}

Intuitively, a deviation $\delta_R>0$ of the range $R_\mathrm{w}$ (higher proton energy) results in a higher number of detected counts due to an increased region of \gr production and a larger subtended solid angle for points closer to the detector. A third, minor reason is the shorter attenuation path for \grs emitted closer to the detector. To disentangle the two main contributions, two further partial derivatives are calculated.

For the contribution of the subtended solid angle, we study the relative count variation $\Delta_\Omega$ for an error $\delta_z$ in the phantom position $z_\mathrm{0}$, while keeping the same proton range and detector position $z_\mathrm{d}$. For symmetry, this analysis is comparable to the study of an error $-\delta_z$ in the detector position $z_\mathrm{d}$ with a fixed phantom position $z_0$:
\begin{eqnarray}
\Delta_{\Omega}& \equiv &-\frac{1}{\hat{N}_\mathrm{d}}\frac{\partial\hat{N}_\mathrm{d}}{\partial z_\mathrm{d}}(z_\mathrm{d})\cdot\delta_z \label{eq:gamma_omega_dif}\\
& = &\frac{\delta_z\eta}{2\hat{N}_\mathrm{d}}\int\limits_{z_0}^{z_0+L} \frac{\dif N_\gamma}{\dif z}\frac{\tau r^2 \dif z}{(z_\mathrm{d}-z)^3}\cdot\left(1+\frac{r^2}{(z_\mathrm{d}-z)^2}\right)^{-1.5} \label{eq:deltaO}
\end{eqnarray}

For the second effect, namely the increase in the region of prompt \gr production, we analyze the relative variation $\Delta_\gamma$ in the integral \gr emission $N_\gamma$, \cf \refeq{eq:gammaint}, for an error $\delta_R$ in the proton range $R_\mathrm{w}$:
\begin{eqnarray}
\Delta_{\gamma}& \equiv &\frac{1}{N_\gamma}\frac{\partial N_\gamma}{\partial R_\mathrm{w}}(R_\mathrm{w})\cdot\delta_R \label{eq:gamma_emission_dif}\\
& \approx & \frac{\delta_R}{R_\mathrm{w}} \label{eq:deltaG}
\end{eqnarray}

A more clinically realistic scenario is studied: the introduction of an air cavity of thickness $\kappa=2\,$mm within the homogeneous water target (length $L$ is unchanged), \cf \refig{fig:setup}, with its front face located at a position $z_\kappa<z_0+R_\mathrm{w}$, where $z_\kappa-z_0+\kappa<L$. Neglecting the \gr production in air, the prompt \gr detection $\hat{N}_\mathrm{d,\kappa}$ per unit proton in this heterogeneous geometry is derived from:
\begin{multline}
\hat{N}_\mathrm{d;\kappa}=\int\limits_{z_0}^{z_0+L}\frac{\Omega}{4\pi}(z;z_\mathrm{d})\cdot\eta \cdot\dif z \cdot\\\cdot
\begin{cases} 
       \od{N_\gamma}{z}(z;R_\mathrm{w})\cdot\tau(z+\kappa) &, z < z_\kappa \\
       0                                                   &, z\in[z_\kappa ,z_\kappa+\kappa] \\
       \od{N_\gamma}{z}(z;R_\mathrm{w}+\kappa)\cdot\tau(z) &, z > z_\kappa+\kappa
\end{cases}
\label{eq:nkappa}
\end{multline}
and the relative difference in counts with respect to the nominal target follows from:
\begin{equation}
\Delta_\kappa = \hat{N}_\mathrm{d;\kappa}/\hat{N}_\mathrm{d;0} - 1
\label{eq:deltaK}
\end{equation}

In any of the presented scenarios, the statistical precision of the measurement is essential to detect small range errors. According to Poisson statistics and including partial suppression of background events based on energy and timing filters \cite{cambraia_background}, the relative uncertainty $\Delta_\mathrm{s}$ yields:
\begin{equation}
\Delta_\mathrm{s}=\frac{1}{\sqrt{N_\mathrm{d}}}\cdot\sqrt{1+(1-\chi)\cdot\frac{\Sigma_\mathrm{t}}{T_\mathrm{p}}\cdot(k_\gamma-1)}
\label{eq:deltaS}
\end{equation}
where $\chi\approx0.5$ is the fraction of background events that can be separated from prompt \grs based on energy filters, $\Sigma_\mathrm{t}\approx2$\,ns is the proton bunch time \FWHM measured for the C230 cyclotron \cite{petzoldt_pbm}, and $T_\mathrm{p}=9.4$\,ns is the proton bunch period, \ie the inverse of the cyclotron \RF.

\begin{table}
\begin{tabular}{lll}
$z$ & beam axis coordinate & [mm]\\
$z_0$ & front face position of water target & 0\,mm\\
$z_\mathrm{d}$ & front face position of the detector & 152\,mm\\
$L$ & length of the water phantom & 150\,mm\\
$R_\mathrm{w}$ & proton range in water & [mm]\\
$r$ & detector radius & \nicefrac{1}{2}''\\
$t$ & detector thickness & 1\nicefrac{1}{2}''\\

$N_\gamma$ & \grs emitted per incident proton & [ ]\\
$\bar{g}_\lambda$ & prompt \gr yield per unit length & 6$\cdot$10\tup{-4}\,mm\tup{-1}\\
$\bar{f}$ & distal falloff slope & 0.3\,mm\tup{-1}\\

$\tau$ & transmission probability  & [ ]\\
$\bar{\mu}_\mathrm{w}$ & average absorption coefficient of water & 10\tup{-4}\,mm\tup{-1}\\

$\Omega/4\pi$ & relative solid angle subtended & [ ]\\
$\eta$ & probability of interaction in detector & [ ]\\
$\bar{\mu}_\mathrm{d}$ & average attenuation coefficient of detector & 2$\cdot$10\tup{-2}\,mm\tup{-1}\\
$t_\mathrm{d}$ & decay time of the scintillation light pulse & 16\,ns\\

$N_\mathrm{p}$ & number of protons within one spot cluster & 10\tup{8}\\
$\hat{N}_\mathrm{d}$ & prompt \grs detected per proton (theory)& [ ]\\
$N_\mathrm{d}$ & prompt \grs detected per spot (theory) & [ ]\\
$M_\mathrm{d}$ & prompt \grs measured per spot cluster & [ ]\\

$k_\gamma$ & ratio between all counts and the prompt \grs & 2.5  \\

$\dot{N}_\mathrm{d,all}$ & overall detector count rate & [Mcps]\\
$I_\mathrm{p}$ & the proton beam (peak) current at the target & 2\,nA\\
$e$ & elementary charge & 1.6$\cdot$10\tup{-19}\,C \\

$\delta_R$ & absolute error in the proton range $R_\mathrm{w}$ & 1\,mm\\
$\delta_z$ & absolute error in the phantom position $z_0$ & 1\,mm\\
$\kappa$ & thickness of a cavity within the water target & 2\,mm\\
$z_\kappa$ & front face position of a cavity within the target & [mm]\\
$\Delta_N$ & relative count variation for a range error $\delta_R$ & [ ]\\
$\Delta_\Omega$ & relative count variation for a target error $\delta_z$ & [ ]\\
$\Delta_\gamma$ & \gr emission variation for a range error $\delta_R$ & [ ]\\
$\Delta_\kappa$ & relative count variation due to a cavity $\kappa$ & [ ]\\
$\Delta_\mathrm{s}$ & statistical measurement uncertainty & [ ]\\

$\chi$ & background fraction separated by energy filters & 0.5\\
$\Sigma_\mathrm{t}$ & proton bunch time spread (FWHM) & 2\,ns\\
$T_\mathrm{p}$ & proton bunch period & 9.4\,ns
 
\end{tabular}
\vspace{1mm}
\caption{Variables and parameters used within the theoretical model of coaxial prompt \gr detection, \cf \refig{fig:setup}.}
\label{tab:params}
\end{table}

\subsection{Validation}\label{ss:validation}
The developed analytical model is compared with \MC simulations by means of the TOPAS software \cite{TOPAS12}, version 3.2, which is based on the Geant4 10.05.p01 toolkit \cite{ALLISON2016186}.
The proton beam has a lateral spread of $5$\,mm, no energy or angular spread, and propagates towards the water phantom. The proton beam kinetic energy is varied between $1$ and $125$\,MeV in steps of $1$\,MeV, and is correlated to the \ac{CSDA} proton range $R_\mathrm{w}$ in water via \cite{nist_estar}. For every beam energy, $N_\mathrm{p}=10^8$ histories are generated. The default physics lists are deployed, and a cutoff of $0.5\,$mm is set for all particles.

The water target is a cylinder with $200\,$mm diameter and $L=150$\,mm length. The environment surrounding the target is set to air.
The detector is modeled as a cylinder made of LaBr\tlow{3} and is located as described in \refig{fig:setup} and table~\ref{tab:params}.

Two different energy deposit scorers (with $1000$ bins between $0$ and $10$\,MeV) are defined in the detector volume. The first one scores particles (mainly photons) stemming from a neutron or positron interaction (background), \eg \grs due to neutron capture on hydrogen; the second one records the rest (signal), \ie prompt \grs. A detector sensitivity threshold of $50$\,keV is applied in the posterior analysis. A third scorer is defined in order to track the total number of prompt \grs emitted across the beam path.

For a beam energy of $100$\,MeV, a dedicated phase space scorer of emitted prompt \grs (stemming from proton inelastic reactions) is defined, in order to obtain a differential (energy-integrated) emission profile $\dif N_\gamma / \dif z$ along the $z$ axis, to be compared against \refeq{eq:dNe}.
 
\section{Results}\label{sec:results}

In \refig{fig:gammaRate}, the overall detector count rate $\dot{N}_\mathrm{d,all}$ as a function of the proton range $R_\mathrm{w}$ is calculated for the proposed coaxial detection setup, \cf \refig{fig:setup} and table~\ref{tab:params}. As expected, the higher the incoming proton energy (penetration depth), the larger the count rate in the detector due to an increased \gr production, solid angle subtended, and decreased attenuation path. The dashed lines indicate the virtual count rate in case of a target alignment error of $\pm\,\text{2}\,$mm (across the beam axis). The red data points correspond to selected proton ranges (beam energy layers). Their horizontal error bar is set to 0.5\,mm (commissioning tolerance), while the vertical error represents the predicted statistical uncertainty $\Delta_\mathrm{s}\times\dot{N}_\mathrm{d,all}$ assuming $N_\mathrm{p}=\text{10}^\text{8}$ protons per spot cluster.

\begin{figure}[h]
\centering
\scalebox{0.4}{

 }
\caption{Modeled \gr count rate $\dot{N}_\mathrm{d,all}$ as a function of proton range $R_\mathrm{w}$, \cf \refeqs{eq:gamma_counts} and \ref{eq:dload}, of a $\diameter$1''$\times$1\nicefrac{1}{2}'' LaBr\tlow{3} detector located on the beam axis behind a 150\,mm long water phantom, as described in \refig{fig:setup} and table~\ref{tab:params}. The proton beam current at the target is $I_\mathrm{p}=\text{2}$\,nA and the number of protons per spot cluster is $N_\mathrm{p}=$\,10\tup{8}. The solid line refers to the nominal situation, whereas the dashed lines correspond to the results expected for a virtual target misalignment of $\pm\,\text{2}\,$mm (along the beam axis). The red points are selected proton ranges accompanied with 0.5\,mm horizontal error bars, and with vertical error bars that represent the statistical uncertainty $\Delta_\mathrm{s}\times\dot{N}_\mathrm{d,all}$ according to the collected prompt \gr counts per spot, \cf \refeqs{eq:spot_counts} and \ref{eq:deltaS}.}
\label{fig:gammaRate}
\end{figure}

The relative variation of the solid curve in \refig{fig:gammaRate} is depicted in \refig{fig:gammaDeltaN} as a function of proton range $R_\mathrm{w}$, \cf \refeq{eq:deltaN}. In this case, the statistical uncertainty $\Delta_\mathrm{s}$ is superimposed as a dashed line, and imposes a theoretical precision limit for a spot cluster with $N_\mathrm{p}=\text{10}^\text{8}$ protons. The concave shape of the solid curve is mainly due to the interplay of two effects: the slower (relative) rise in \gr production with proton range for deeper irradiation points and the faster increase of the solid angle subtended by the detector.

Both contributions are isolated in \refigs{fig:gammaDeltaG} and \ref{fig:gammaDeltaO}, respectively, by calculating the partial derivatives of \refeqs{eq:deltaG} and \ref{eq:deltaO}. The combination of both effects leads to the shape visible in \refig{fig:gammaDeltaN}. It should be noted that, in the case of \refig{fig:gammaDeltaO}, a target misalignment of 1\,mm would not be measurable for a proton range below 60\,mm. The reason is that the statistical measurement uncertainty, considering 10\tup{8} delivered protons, is larger than the expected change in the prompt \gr signal.

\begin{figure}[h]
\hspace{6mm}\scalebox{0.4}{\begin{tikzpicture}
\pgfdeclareplotmark{cross} {
\pgfpathmoveto{\pgfpoint{-0.3\pgfplotmarksize}{\pgfplotmarksize}}
\pgfpathlineto{\pgfpoint{+0.3\pgfplotmarksize}{\pgfplotmarksize}}
\pgfpathlineto{\pgfpoint{+0.3\pgfplotmarksize}{0.3\pgfplotmarksize}}
\pgfpathlineto{\pgfpoint{+1\pgfplotmarksize}{0.3\pgfplotmarksize}}
\pgfpathlineto{\pgfpoint{+1\pgfplotmarksize}{-0.3\pgfplotmarksize}}
\pgfpathlineto{\pgfpoint{+0.3\pgfplotmarksize}{-0.3\pgfplotmarksize}}
\pgfpathlineto{\pgfpoint{+0.3\pgfplotmarksize}{-1.\pgfplotmarksize}}
\pgfpathlineto{\pgfpoint{-0.3\pgfplotmarksize}{-1.\pgfplotmarksize}}
\pgfpathlineto{\pgfpoint{-0.3\pgfplotmarksize}{-0.3\pgfplotmarksize}}
\pgfpathlineto{\pgfpoint{-1.\pgfplotmarksize}{-0.3\pgfplotmarksize}}
\pgfpathlineto{\pgfpoint{-1.\pgfplotmarksize}{0.3\pgfplotmarksize}}
\pgfpathlineto{\pgfpoint{-0.3\pgfplotmarksize}{0.3\pgfplotmarksize}}
\pgfpathclose
\pgfusepathqstroke
}
\pgfdeclareplotmark{cross*} {
\pgfpathmoveto{\pgfpoint{-0.3\pgfplotmarksize}{\pgfplotmarksize}}
\pgfpathlineto{\pgfpoint{+0.3\pgfplotmarksize}{\pgfplotmarksize}}
\pgfpathlineto{\pgfpoint{+0.3\pgfplotmarksize}{0.3\pgfplotmarksize}}
\pgfpathlineto{\pgfpoint{+1\pgfplotmarksize}{0.3\pgfplotmarksize}}
\pgfpathlineto{\pgfpoint{+1\pgfplotmarksize}{-0.3\pgfplotmarksize}}
\pgfpathlineto{\pgfpoint{+0.3\pgfplotmarksize}{-0.3\pgfplotmarksize}}
\pgfpathlineto{\pgfpoint{+0.3\pgfplotmarksize}{-1.\pgfplotmarksize}}
\pgfpathlineto{\pgfpoint{-0.3\pgfplotmarksize}{-1.\pgfplotmarksize}}
\pgfpathlineto{\pgfpoint{-0.3\pgfplotmarksize}{-0.3\pgfplotmarksize}}
\pgfpathlineto{\pgfpoint{-1.\pgfplotmarksize}{-0.3\pgfplotmarksize}}
\pgfpathlineto{\pgfpoint{-1.\pgfplotmarksize}{0.3\pgfplotmarksize}}
\pgfpathlineto{\pgfpoint{-0.3\pgfplotmarksize}{0.3\pgfplotmarksize}}
\pgfpathclose
\pgfusepathqfillstroke
}
\pgfdeclareplotmark{newstar} {
\pgfpathmoveto{\pgfqpoint{0pt}{\pgfplotmarksize}}
\pgfpathlineto{\pgfqpointpolar{44}{0.5\pgfplotmarksize}}
\pgfpathlineto{\pgfqpointpolar{18}{\pgfplotmarksize}}
\pgfpathlineto{\pgfqpointpolar{-20}{0.5\pgfplotmarksize}}
\pgfpathlineto{\pgfqpointpolar{-54}{\pgfplotmarksize}}
\pgfpathlineto{\pgfqpointpolar{-90}{0.5\pgfplotmarksize}}
\pgfpathlineto{\pgfqpointpolar{234}{\pgfplotmarksize}}
\pgfpathlineto{\pgfqpointpolar{198}{0.5\pgfplotmarksize}}
\pgfpathlineto{\pgfqpointpolar{162}{\pgfplotmarksize}}
\pgfpathlineto{\pgfqpointpolar{134}{0.5\pgfplotmarksize}}
\pgfpathclose
\pgfusepathqstroke
}
\pgfdeclareplotmark{newstar*} {
\pgfpathmoveto{\pgfqpoint{0pt}{\pgfplotmarksize}}
\pgfpathlineto{\pgfqpointpolar{44}{0.5\pgfplotmarksize}}
\pgfpathlineto{\pgfqpointpolar{18}{\pgfplotmarksize}}
\pgfpathlineto{\pgfqpointpolar{-20}{0.5\pgfplotmarksize}}
\pgfpathlineto{\pgfqpointpolar{-54}{\pgfplotmarksize}}
\pgfpathlineto{\pgfqpointpolar{-90}{0.5\pgfplotmarksize}}
\pgfpathlineto{\pgfqpointpolar{234}{\pgfplotmarksize}}
\pgfpathlineto{\pgfqpointpolar{198}{0.5\pgfplotmarksize}}
\pgfpathlineto{\pgfqpointpolar{162}{\pgfplotmarksize}}
\pgfpathlineto{\pgfqpointpolar{134}{0.5\pgfplotmarksize}}
\pgfpathclose
\pgfusepathqfillstroke
}
\definecolor{c}{rgb}{0,0,0};
\draw [c,line width=0.9] (2,1.5) -- (2,13.5) -- (18,13.5) -- (18,1.5) -- (2,1.5);
\draw [c,line width=0.9] (2,1.5) -- (2,13.5) -- (18,13.5) -- (18,1.5) -- (2,1.5);
\draw [c,line width=0.9] (2,1.5) -- (2,13.5) -- (18,13.5) -- (18,1.5) -- (2,1.5);
\draw [c,line width=0.9] (2,1.5) -- (2,13.5) -- (18,13.5) -- (18,1.5) -- (2,1.5);
\draw [c,line width=0.9] (2,1.5) -- (18,1.5);
\draw [c,dash pattern=on 0.80pt off 1.60pt ,line width=0.9] (2,13.5) -- (2,1.5);
\draw [c,dash pattern=on 0.80pt off 1.60pt ,line width=0.9] (3.6,13.5) -- (3.6,1.5);
\draw [c,dash pattern=on 0.80pt off 1.60pt ,line width=0.9] (5.2,13.5) -- (5.2,1.5);
\draw [c,dash pattern=on 0.80pt off 1.60pt ,line width=0.9] (6.8,13.5) -- (6.8,1.5);
\draw [c,dash pattern=on 0.80pt off 1.60pt ,line width=0.9] (8.4,13.5) -- (8.4,1.5);
\draw [c,dash pattern=on 0.80pt off 1.60pt ,line width=0.9] (10,13.5) -- (10,1.5);
\draw [c,dash pattern=on 0.80pt off 1.60pt ,line width=0.9] (11.6,13.5) -- (11.6,1.5);
\draw [c,dash pattern=on 0.80pt off 1.60pt ,line width=0.9] (13.2,13.5) -- (13.2,1.5);
\draw [c,dash pattern=on 0.80pt off 1.60pt ,line width=0.9] (14.8,13.5) -- (14.8,1.5);
\draw [c,dash pattern=on 0.80pt off 1.60pt ,line width=0.9] (16.4,13.5) -- (16.4,1.5);
\draw [c,dash pattern=on 0.80pt off 1.60pt ,line width=0.9] (18,13.5) -- (18,1.5);
\draw [c,line width=0.9] (2,1.5) -- (2,13.5);
\draw [c,dash pattern=on 0.80pt off 1.60pt ,line width=0.9] (18,1.5) -- (2,1.5);
\draw [c,dash pattern=on 0.80pt off 1.60pt ,line width=0.9] (18,3.5) -- (2,3.5);
\draw [c,dash pattern=on 0.80pt off 1.60pt ,line width=0.9] (18,5.5) -- (2,5.5);
\draw [c,dash pattern=on 0.80pt off 1.60pt ,line width=0.9] (18,7.5) -- (2,7.5);
\draw [c,dash pattern=on 0.80pt off 1.60pt ,line width=0.9] (18,9.5) -- (2,9.5);
\draw [c,dash pattern=on 0.80pt off 1.60pt ,line width=0.9] (18,11.5) -- (2,11.5);
\draw [c,dash pattern=on 0.80pt off 1.60pt ,line width=0.9] (18,13.5) -- (2,13.5);
\draw [c,line width=0.9] (2,1.5) -- (18,1.5);
\draw [c,line width=0.9] (2,1.86) -- (2,1.5);
\draw [c,line width=0.9] (2.32,1.68) -- (2.32,1.5);
\draw [c,line width=0.9] (2.64,1.68) -- (2.64,1.5);
\draw [c,line width=0.9] (2.96,1.68) -- (2.96,1.5);
\draw [c,line width=0.9] (3.28,1.68) -- (3.28,1.5);
\draw [c,line width=0.9] (3.6,1.86) -- (3.6,1.5);
\draw [c,line width=0.9] (3.92,1.68) -- (3.92,1.5);
\draw [c,line width=0.9] (4.24,1.68) -- (4.24,1.5);
\draw [c,line width=0.9] (4.56,1.68) -- (4.56,1.5);
\draw [c,line width=0.9] (4.88,1.68) -- (4.88,1.5);
\draw [c,line width=0.9] (5.2,1.86) -- (5.2,1.5);
\draw [c,line width=0.9] (5.52,1.68) -- (5.52,1.5);
\draw [c,line width=0.9] (5.84,1.68) -- (5.84,1.5);
\draw [c,line width=0.9] (6.16,1.68) -- (6.16,1.5);
\draw [c,line width=0.9] (6.48,1.68) -- (6.48,1.5);
\draw [c,line width=0.9] (6.8,1.86) -- (6.8,1.5);
\draw [c,line width=0.9] (7.12,1.68) -- (7.12,1.5);
\draw [c,line width=0.9] (7.44,1.68) -- (7.44,1.5);
\draw [c,line width=0.9] (7.76,1.68) -- (7.76,1.5);
\draw [c,line width=0.9] (8.08,1.68) -- (8.08,1.5);
\draw [c,line width=0.9] (8.4,1.86) -- (8.4,1.5);
\draw [c,line width=0.9] (8.72,1.68) -- (8.72,1.5);
\draw [c,line width=0.9] (9.04,1.68) -- (9.04,1.5);
\draw [c,line width=0.9] (9.36,1.68) -- (9.36,1.5);
\draw [c,line width=0.9] (9.68,1.68) -- (9.68,1.5);
\draw [c,line width=0.9] (10,1.86) -- (10,1.5);
\draw [c,line width=0.9] (10.32,1.68) -- (10.32,1.5);
\draw [c,line width=0.9] (10.64,1.68) -- (10.64,1.5);
\draw [c,line width=0.9] (10.96,1.68) -- (10.96,1.5);
\draw [c,line width=0.9] (11.28,1.68) -- (11.28,1.5);
\draw [c,line width=0.9] (11.6,1.86) -- (11.6,1.5);
\draw [c,line width=0.9] (11.92,1.68) -- (11.92,1.5);
\draw [c,line width=0.9] (12.24,1.68) -- (12.24,1.5);
\draw [c,line width=0.9] (12.56,1.68) -- (12.56,1.5);
\draw [c,line width=0.9] (12.88,1.68) -- (12.88,1.5);
\draw [c,line width=0.9] (13.2,1.86) -- (13.2,1.5);
\draw [c,line width=0.9] (13.52,1.68) -- (13.52,1.5);
\draw [c,line width=0.9] (13.84,1.68) -- (13.84,1.5);
\draw [c,line width=0.9] (14.16,1.68) -- (14.16,1.5);
\draw [c,line width=0.9] (14.48,1.68) -- (14.48,1.5);
\draw [c,line width=0.9] (14.8,1.86) -- (14.8,1.5);
\draw [c,line width=0.9] (15.12,1.68) -- (15.12,1.5);
\draw [c,line width=0.9] (15.44,1.68) -- (15.44,1.5);
\draw [c,line width=0.9] (15.76,1.68) -- (15.76,1.5);
\draw [c,line width=0.9] (16.08,1.68) -- (16.08,1.5);
\draw [c,line width=0.9] (16.4,1.86) -- (16.4,1.5);
\draw [c,line width=0.9] (16.72,1.68) -- (16.72,1.5);
\draw [c,line width=0.9] (17.04,1.68) -- (17.04,1.5);
\draw [c,line width=0.9] (17.36,1.68) -- (17.36,1.5);
\draw [c,line width=0.9] (17.68,1.68) -- (17.68,1.5);
\draw [c,line width=0.9] (18,1.86) -- (18,1.5);
\draw [anchor=base] (2,0.825) node[scale=1.66573, color=c, rotate=0]{20};
\draw [anchor=base] (3.6,0.825) node[scale=1.66573, color=c, rotate=0]{30};
\draw [anchor=base] (5.2,0.825) node[scale=1.66573, color=c, rotate=0]{40};
\draw [anchor=base] (6.8,0.825) node[scale=1.66573, color=c, rotate=0]{50};
\draw [anchor=base] (8.4,0.825) node[scale=1.66573, color=c, rotate=0]{60};
\draw [anchor=base] (10,0.825) node[scale=1.66573, color=c, rotate=0]{70};
\draw [anchor=base] (11.6,0.825) node[scale=1.66573, color=c, rotate=0]{80};
\draw [anchor=base] (13.2,0.825) node[scale=1.66573, color=c, rotate=0]{90};
\draw [anchor=base] (14.8,0.825) node[scale=1.66573, color=c, rotate=0]{100};
\draw [anchor=base] (16.4,0.825) node[scale=1.66573, color=c, rotate=0]{110};
\draw [anchor=base] (18,0.825) node[scale=1.66573, color=c, rotate=0]{120};
\draw [anchor= east] (18,0.3) node[scale=1.66573, color=c, rotate=0]{$\text{Proton range}\,\,R_\mathrm{w}\,/\,\text{mm}$};
\draw [c,line width=0.9] (2,1.5) -- (2,13.5);
\draw [c,line width=0.9] (2.48,1.5) -- (2,1.5);
\draw [c,line width=0.9] (2.24,1.9) -- (2,1.9);
\draw [c,line width=0.9] (2.24,2.3) -- (2,2.3);
\draw [c,line width=0.9] (2.24,2.7) -- (2,2.7);
\draw [c,line width=0.9] (2.24,3.1) -- (2,3.1);
\draw [c,line width=0.9] (2.48,3.5) -- (2,3.5);
\draw [c,line width=0.9] (2.24,3.9) -- (2,3.9);
\draw [c,line width=0.9] (2.24,4.3) -- (2,4.3);
\draw [c,line width=0.9] (2.24,4.7) -- (2,4.7);
\draw [c,line width=0.9] (2.24,5.1) -- (2,5.1);
\draw [c,line width=0.9] (2.48,5.5) -- (2,5.5);
\draw [c,line width=0.9] (2.24,5.9) -- (2,5.9);
\draw [c,line width=0.9] (2.24,6.3) -- (2,6.3);
\draw [c,line width=0.9] (2.24,6.7) -- (2,6.7);
\draw [c,line width=0.9] (2.24,7.1) -- (2,7.1);
\draw [c,line width=0.9] (2.48,7.5) -- (2,7.5);
\draw [c,line width=0.9] (2.24,7.9) -- (2,7.9);
\draw [c,line width=0.9] (2.24,8.3) -- (2,8.3);
\draw [c,line width=0.9] (2.24,8.7) -- (2,8.7);
\draw [c,line width=0.9] (2.24,9.1) -- (2,9.1);
\draw [c,line width=0.9] (2.48,9.5) -- (2,9.5);
\draw [c,line width=0.9] (2.24,9.9) -- (2,9.9);
\draw [c,line width=0.9] (2.24,10.3) -- (2,10.3);
\draw [c,line width=0.9] (2.24,10.7) -- (2,10.7);
\draw [c,line width=0.9] (2.24,11.1) -- (2,11.1);
\draw [c,line width=0.9] (2.48,11.5) -- (2,11.5);
\draw [c,line width=0.9] (2.24,11.9) -- (2,11.9);
\draw [c,line width=0.9] (2.24,12.3) -- (2,12.3);
\draw [c,line width=0.9] (2.24,12.7) -- (2,12.7);
\draw [c,line width=0.9] (2.24,13.1) -- (2,13.1);
\draw [c,line width=0.9] (2.48,13.5) -- (2,13.5);
\draw [anchor= east] (1.9,1.5) node[scale=1.66573, color=c, rotate=0]{0};
\draw [anchor= east] (1.9,3.5) node[scale=1.66573, color=c, rotate=0]{1};
\draw [anchor= east] (1.9,5.5) node[scale=1.66573, color=c, rotate=0]{2};
\draw [anchor= east] (1.9,7.5) node[scale=1.66573, color=c, rotate=0]{3};
\draw [anchor= east] (1.9,9.5) node[scale=1.66573, color=c, rotate=0]{4};
\draw [anchor= east] (1.9,11.5) node[scale=1.66573, color=c, rotate=0]{5};
\draw [anchor= east] (1.9,13.5) node[scale=1.66573, color=c, rotate=0]{6};
\draw [anchor= east] (0.633333,13.5) node[scale=1.66573, color=c, rotate=90]{$\text{Relative count variation}\,\,\Delta_{N}(\delta_{R}=1\,\text{mm})\,/\,\,\\\%$};
\draw [c,line width=1.8] (2.8,11.0544) -- (2.96,10.7599) -- (3.12,10.4883) -- (3.28,10.2372) -- (3.44,10.0044) -- (3.6,9.7882) -- (3.76,9.58698) -- (3.92,9.39938) -- (4.08,9.22417) -- (4.24,9.0603) -- (4.4,8.90681) -- (4.56,8.76288) -- (4.72,8.62774)
 -- (4.88,8.50073) -- (5.04,8.38126) -- (5.2,8.26879) -- (5.36,8.16282) -- (5.52,8.06293) -- (5.68,7.96871) -- (5.84,7.87981) -- (6,7.7959) -- (6.16,7.71669) -- (6.32,7.64189) -- (6.48,7.57128) -- (6.64,7.50461) -- (6.8,7.44169) -- (6.96,7.38232) --
 (7.12,7.32633) -- (7.28,7.27356) -- (7.44,7.22387) -- (7.6,7.17712) -- (7.76,7.13318) -- (7.92,7.09195) -- (8.08,7.05331) -- (8.24,7.01718) -- (8.4,6.98346) -- (8.56,6.95208) -- (8.72,6.92296) -- (8.88,6.89603) -- (9.04,6.87123) -- (9.2,6.84851) --
 (9.36,6.82781) -- (9.52,6.80909) -- (9.68,6.79231) -- (9.84,6.77743) -- (10,6.76442) -- (10.16,6.75325) -- (10.32,6.7439) -- (10.48,6.73634) -- (10.64,6.73057) -- (10.8,6.72656) -- (10.96,6.7243) -- (11.12,6.72379) -- (11.28,6.72504) --
 (11.44,6.72803) -- (11.6,6.73277) -- (11.76,6.73927) -- (11.92,6.74754) -- (12.08,6.7576) -- (12.24,6.76946) -- (12.4,6.78314) -- (12.56,6.79868) -- (12.72,6.81609) -- (12.88,6.83542) -- (13.04,6.85669) -- (13.2,6.87996) -- (13.36,6.90526) --
 (13.52,6.93265) -- (13.68,6.96218) -- (13.84,6.99392) -- (14,7.02793) -- (14.16,7.06427) -- (14.32,7.10304) -- (14.48,7.14431) -- (14.64,7.18818) -- (14.8,7.23475) -- (14.96,7.28412) -- (15.12,7.33641) -- (15.28,7.39174) -- (15.44,7.45025) --
 (15.6,7.51209) -- (15.76,7.57742) -- (15.92,7.6464) -- (16.08,7.71921) -- (16.24,7.79607) -- (16.4,7.87718) -- (16.56,7.96277) -- (16.72,8.05309) -- (16.88,8.14842) -- (17.04,8.24905) -- (17.2,8.3553);
\draw [c,dash pattern=on 2.40pt off 2.40pt ,line width=0.9] (2.8,6.7116) -- (2.96,6.59049) -- (3.12,6.47571) -- (3.28,6.3667) -- (3.44,6.26295) -- (3.6,6.16403) -- (3.76,6.06954) -- (3.92,5.97915) -- (4.08,5.89253) -- (4.24,5.80941) -- (4.4,5.72955)
 -- (4.56,5.6527) -- (4.72,5.57868) -- (4.88,5.50728) -- (5.04,5.43835) -- (5.2,5.37173) -- (5.36,5.30727) -- (5.52,5.24485) -- (5.68,5.18434) -- (5.84,5.12565) -- (6,5.06866) -- (6.16,5.01328) -- (6.32,4.95942) -- (6.48,4.90701) -- (6.64,4.85597) --
 (6.8,4.80623) -- (6.96,4.75773) -- (7.12,4.7104) -- (7.28,4.66418) -- (7.44,4.61903) -- (7.6,4.5749) -- (7.76,4.53174) -- (7.92,4.48949) -- (8.08,4.44814) -- (8.24,4.40762) -- (8.4,4.36791) -- (8.56,4.32898) -- (8.72,4.29078) -- (8.88,4.2533) --
 (9.04,4.21649) -- (9.2,4.18033) -- (9.36,4.1448) -- (9.52,4.10987) -- (9.68,4.07551) -- (9.84,4.04171) -- (10,4.00843) -- (10.16,3.97567) -- (10.32,3.9434) -- (10.48,3.9116) -- (10.64,3.88025) -- (10.8,3.84934) -- (10.96,3.81885) -- (11.12,3.78877)
 -- (11.28,3.75907) -- (11.44,3.72974) -- (11.6,3.70078) -- (11.76,3.67216) -- (11.92,3.64387) -- (12.08,3.6159) -- (12.24,3.58824) -- (12.4,3.56088) -- (12.56,3.5338) -- (12.72,3.50699) -- (12.88,3.48045) -- (13.04,3.45415) -- (13.2,3.4281) --
 (13.36,3.40228) -- (13.52,3.37669) -- (13.68,3.3513) -- (13.84,3.32612) -- (14,3.30114) -- (14.16,3.27634) -- (14.32,3.25171) -- (14.48,3.22726) -- (14.64,3.20296) -- (14.8,3.17882) -- (14.96,3.15482) -- (15.12,3.13096) -- (15.28,3.10722) --
 (15.44,3.08361) -- (15.6,3.06011) -- (15.76,3.03671) -- (15.92,3.01341) -- (16.08,2.9902) -- (16.24,2.96706) -- (16.4,2.94401) -- (16.56,2.92102) -- (16.72,2.89809) -- (16.88,2.87521) -- (17.04,2.85237) -- (17.2,2.82957);
\end{tikzpicture}
 }
\vspace{-12pt}
\caption{Solid line: Relative change $\Delta_N$ of measured \grs as a function of proton range $R_\mathrm{w}$ for an error $\delta_R=\text{1}\,\text{mm}$, \cf \refeq{eq:deltaN}, for the coaxial detection setup described in \refig{fig:setup} and table~\ref{tab:params}. Dashed line: Statistical measurement precision $\Delta_\mathrm{s}$, \cf \refeq{eq:deltaS}, for a spot cluster with $N_\mathrm{p}=\text{10}^\text{8}$ protons.}
\label{fig:gammaDeltaN}

\vspace{8pt}
\hspace{6mm}\scalebox{0.4}{\begin{tikzpicture}
\pgfdeclareplotmark{cross} {
\pgfpathmoveto{\pgfpoint{-0.3\pgfplotmarksize}{\pgfplotmarksize}}
\pgfpathlineto{\pgfpoint{+0.3\pgfplotmarksize}{\pgfplotmarksize}}
\pgfpathlineto{\pgfpoint{+0.3\pgfplotmarksize}{0.3\pgfplotmarksize}}
\pgfpathlineto{\pgfpoint{+1\pgfplotmarksize}{0.3\pgfplotmarksize}}
\pgfpathlineto{\pgfpoint{+1\pgfplotmarksize}{-0.3\pgfplotmarksize}}
\pgfpathlineto{\pgfpoint{+0.3\pgfplotmarksize}{-0.3\pgfplotmarksize}}
\pgfpathlineto{\pgfpoint{+0.3\pgfplotmarksize}{-1.\pgfplotmarksize}}
\pgfpathlineto{\pgfpoint{-0.3\pgfplotmarksize}{-1.\pgfplotmarksize}}
\pgfpathlineto{\pgfpoint{-0.3\pgfplotmarksize}{-0.3\pgfplotmarksize}}
\pgfpathlineto{\pgfpoint{-1.\pgfplotmarksize}{-0.3\pgfplotmarksize}}
\pgfpathlineto{\pgfpoint{-1.\pgfplotmarksize}{0.3\pgfplotmarksize}}
\pgfpathlineto{\pgfpoint{-0.3\pgfplotmarksize}{0.3\pgfplotmarksize}}
\pgfpathclose
\pgfusepathqstroke
}
\pgfdeclareplotmark{cross*} {
\pgfpathmoveto{\pgfpoint{-0.3\pgfplotmarksize}{\pgfplotmarksize}}
\pgfpathlineto{\pgfpoint{+0.3\pgfplotmarksize}{\pgfplotmarksize}}
\pgfpathlineto{\pgfpoint{+0.3\pgfplotmarksize}{0.3\pgfplotmarksize}}
\pgfpathlineto{\pgfpoint{+1\pgfplotmarksize}{0.3\pgfplotmarksize}}
\pgfpathlineto{\pgfpoint{+1\pgfplotmarksize}{-0.3\pgfplotmarksize}}
\pgfpathlineto{\pgfpoint{+0.3\pgfplotmarksize}{-0.3\pgfplotmarksize}}
\pgfpathlineto{\pgfpoint{+0.3\pgfplotmarksize}{-1.\pgfplotmarksize}}
\pgfpathlineto{\pgfpoint{-0.3\pgfplotmarksize}{-1.\pgfplotmarksize}}
\pgfpathlineto{\pgfpoint{-0.3\pgfplotmarksize}{-0.3\pgfplotmarksize}}
\pgfpathlineto{\pgfpoint{-1.\pgfplotmarksize}{-0.3\pgfplotmarksize}}
\pgfpathlineto{\pgfpoint{-1.\pgfplotmarksize}{0.3\pgfplotmarksize}}
\pgfpathlineto{\pgfpoint{-0.3\pgfplotmarksize}{0.3\pgfplotmarksize}}
\pgfpathclose
\pgfusepathqfillstroke
}
\pgfdeclareplotmark{newstar} {
\pgfpathmoveto{\pgfqpoint{0pt}{\pgfplotmarksize}}
\pgfpathlineto{\pgfqpointpolar{44}{0.5\pgfplotmarksize}}
\pgfpathlineto{\pgfqpointpolar{18}{\pgfplotmarksize}}
\pgfpathlineto{\pgfqpointpolar{-20}{0.5\pgfplotmarksize}}
\pgfpathlineto{\pgfqpointpolar{-54}{\pgfplotmarksize}}
\pgfpathlineto{\pgfqpointpolar{-90}{0.5\pgfplotmarksize}}
\pgfpathlineto{\pgfqpointpolar{234}{\pgfplotmarksize}}
\pgfpathlineto{\pgfqpointpolar{198}{0.5\pgfplotmarksize}}
\pgfpathlineto{\pgfqpointpolar{162}{\pgfplotmarksize}}
\pgfpathlineto{\pgfqpointpolar{134}{0.5\pgfplotmarksize}}
\pgfpathclose
\pgfusepathqstroke
}
\pgfdeclareplotmark{newstar*} {
\pgfpathmoveto{\pgfqpoint{0pt}{\pgfplotmarksize}}
\pgfpathlineto{\pgfqpointpolar{44}{0.5\pgfplotmarksize}}
\pgfpathlineto{\pgfqpointpolar{18}{\pgfplotmarksize}}
\pgfpathlineto{\pgfqpointpolar{-20}{0.5\pgfplotmarksize}}
\pgfpathlineto{\pgfqpointpolar{-54}{\pgfplotmarksize}}
\pgfpathlineto{\pgfqpointpolar{-90}{0.5\pgfplotmarksize}}
\pgfpathlineto{\pgfqpointpolar{234}{\pgfplotmarksize}}
\pgfpathlineto{\pgfqpointpolar{198}{0.5\pgfplotmarksize}}
\pgfpathlineto{\pgfqpointpolar{162}{\pgfplotmarksize}}
\pgfpathlineto{\pgfqpointpolar{134}{0.5\pgfplotmarksize}}
\pgfpathclose
\pgfusepathqfillstroke
}
\definecolor{c}{rgb}{0,0,0};
\draw [c,line width=0.9] (2,1.5) -- (2,13.5) -- (18,13.5) -- (18,1.5) -- (2,1.5);
\draw [c,line width=0.9] (2,1.5) -- (2,13.5) -- (18,13.5) -- (18,1.5) -- (2,1.5);
\draw [c,line width=0.9] (2,1.5) -- (2,13.5) -- (18,13.5) -- (18,1.5) -- (2,1.5);
\draw [c,line width=0.9] (2,1.5) -- (2,13.5) -- (18,13.5) -- (18,1.5) -- (2,1.5);
\draw [c,line width=0.9] (2,1.5) -- (18,1.5);
\draw [c,dash pattern=on 0.80pt off 1.60pt ,line width=0.9] (2,13.5) -- (2,1.5);
\draw [c,dash pattern=on 0.80pt off 1.60pt ,line width=0.9] (3.6,13.5) -- (3.6,1.5);
\draw [c,dash pattern=on 0.80pt off 1.60pt ,line width=0.9] (5.2,13.5) -- (5.2,1.5);
\draw [c,dash pattern=on 0.80pt off 1.60pt ,line width=0.9] (6.8,13.5) -- (6.8,1.5);
\draw [c,dash pattern=on 0.80pt off 1.60pt ,line width=0.9] (8.4,13.5) -- (8.4,1.5);
\draw [c,dash pattern=on 0.80pt off 1.60pt ,line width=0.9] (10,13.5) -- (10,1.5);
\draw [c,dash pattern=on 0.80pt off 1.60pt ,line width=0.9] (11.6,13.5) -- (11.6,1.5);
\draw [c,dash pattern=on 0.80pt off 1.60pt ,line width=0.9] (13.2,13.5) -- (13.2,1.5);
\draw [c,dash pattern=on 0.80pt off 1.60pt ,line width=0.9] (14.8,13.5) -- (14.8,1.5);
\draw [c,dash pattern=on 0.80pt off 1.60pt ,line width=0.9] (16.4,13.5) -- (16.4,1.5);
\draw [c,dash pattern=on 0.80pt off 1.60pt ,line width=0.9] (18,13.5) -- (18,1.5);
\draw [c,line width=0.9] (2,1.5) -- (2,13.5);
\draw [c,dash pattern=on 0.80pt off 1.60pt ,line width=0.9] (18,1.5) -- (2,1.5);
\draw [c,dash pattern=on 0.80pt off 1.60pt ,line width=0.9] (18,3.5) -- (2,3.5);
\draw [c,dash pattern=on 0.80pt off 1.60pt ,line width=0.9] (18,5.5) -- (2,5.5);
\draw [c,dash pattern=on 0.80pt off 1.60pt ,line width=0.9] (18,7.5) -- (2,7.5);
\draw [c,dash pattern=on 0.80pt off 1.60pt ,line width=0.9] (18,9.5) -- (2,9.5);
\draw [c,dash pattern=on 0.80pt off 1.60pt ,line width=0.9] (18,11.5) -- (2,11.5);
\draw [c,dash pattern=on 0.80pt off 1.60pt ,line width=0.9] (18,13.5) -- (2,13.5);
\draw [c,line width=0.9] (2,1.5) -- (18,1.5);
\draw [c,line width=0.9] (2,1.86) -- (2,1.5);
\draw [c,line width=0.9] (2.32,1.68) -- (2.32,1.5);
\draw [c,line width=0.9] (2.64,1.68) -- (2.64,1.5);
\draw [c,line width=0.9] (2.96,1.68) -- (2.96,1.5);
\draw [c,line width=0.9] (3.28,1.68) -- (3.28,1.5);
\draw [c,line width=0.9] (3.6,1.86) -- (3.6,1.5);
\draw [c,line width=0.9] (3.92,1.68) -- (3.92,1.5);
\draw [c,line width=0.9] (4.24,1.68) -- (4.24,1.5);
\draw [c,line width=0.9] (4.56,1.68) -- (4.56,1.5);
\draw [c,line width=0.9] (4.88,1.68) -- (4.88,1.5);
\draw [c,line width=0.9] (5.2,1.86) -- (5.2,1.5);
\draw [c,line width=0.9] (5.52,1.68) -- (5.52,1.5);
\draw [c,line width=0.9] (5.84,1.68) -- (5.84,1.5);
\draw [c,line width=0.9] (6.16,1.68) -- (6.16,1.5);
\draw [c,line width=0.9] (6.48,1.68) -- (6.48,1.5);
\draw [c,line width=0.9] (6.8,1.86) -- (6.8,1.5);
\draw [c,line width=0.9] (7.12,1.68) -- (7.12,1.5);
\draw [c,line width=0.9] (7.44,1.68) -- (7.44,1.5);
\draw [c,line width=0.9] (7.76,1.68) -- (7.76,1.5);
\draw [c,line width=0.9] (8.08,1.68) -- (8.08,1.5);
\draw [c,line width=0.9] (8.4,1.86) -- (8.4,1.5);
\draw [c,line width=0.9] (8.72,1.68) -- (8.72,1.5);
\draw [c,line width=0.9] (9.04,1.68) -- (9.04,1.5);
\draw [c,line width=0.9] (9.36,1.68) -- (9.36,1.5);
\draw [c,line width=0.9] (9.68,1.68) -- (9.68,1.5);
\draw [c,line width=0.9] (10,1.86) -- (10,1.5);
\draw [c,line width=0.9] (10.32,1.68) -- (10.32,1.5);
\draw [c,line width=0.9] (10.64,1.68) -- (10.64,1.5);
\draw [c,line width=0.9] (10.96,1.68) -- (10.96,1.5);
\draw [c,line width=0.9] (11.28,1.68) -- (11.28,1.5);
\draw [c,line width=0.9] (11.6,1.86) -- (11.6,1.5);
\draw [c,line width=0.9] (11.92,1.68) -- (11.92,1.5);
\draw [c,line width=0.9] (12.24,1.68) -- (12.24,1.5);
\draw [c,line width=0.9] (12.56,1.68) -- (12.56,1.5);
\draw [c,line width=0.9] (12.88,1.68) -- (12.88,1.5);
\draw [c,line width=0.9] (13.2,1.86) -- (13.2,1.5);
\draw [c,line width=0.9] (13.52,1.68) -- (13.52,1.5);
\draw [c,line width=0.9] (13.84,1.68) -- (13.84,1.5);
\draw [c,line width=0.9] (14.16,1.68) -- (14.16,1.5);
\draw [c,line width=0.9] (14.48,1.68) -- (14.48,1.5);
\draw [c,line width=0.9] (14.8,1.86) -- (14.8,1.5);
\draw [c,line width=0.9] (15.12,1.68) -- (15.12,1.5);
\draw [c,line width=0.9] (15.44,1.68) -- (15.44,1.5);
\draw [c,line width=0.9] (15.76,1.68) -- (15.76,1.5);
\draw [c,line width=0.9] (16.08,1.68) -- (16.08,1.5);
\draw [c,line width=0.9] (16.4,1.86) -- (16.4,1.5);
\draw [c,line width=0.9] (16.72,1.68) -- (16.72,1.5);
\draw [c,line width=0.9] (17.04,1.68) -- (17.04,1.5);
\draw [c,line width=0.9] (17.36,1.68) -- (17.36,1.5);
\draw [c,line width=0.9] (17.68,1.68) -- (17.68,1.5);
\draw [c,line width=0.9] (18,1.86) -- (18,1.5);
\draw [anchor=base] (2,0.825) node[scale=1.66573, color=c, rotate=0]{20};
\draw [anchor=base] (3.6,0.825) node[scale=1.66573, color=c, rotate=0]{30};
\draw [anchor=base] (5.2,0.825) node[scale=1.66573, color=c, rotate=0]{40};
\draw [anchor=base] (6.8,0.825) node[scale=1.66573, color=c, rotate=0]{50};
\draw [anchor=base] (8.4,0.825) node[scale=1.66573, color=c, rotate=0]{60};
\draw [anchor=base] (10,0.825) node[scale=1.66573, color=c, rotate=0]{70};
\draw [anchor=base] (11.6,0.825) node[scale=1.66573, color=c, rotate=0]{80};
\draw [anchor=base] (13.2,0.825) node[scale=1.66573, color=c, rotate=0]{90};
\draw [anchor=base] (14.8,0.825) node[scale=1.66573, color=c, rotate=0]{100};
\draw [anchor=base] (16.4,0.825) node[scale=1.66573, color=c, rotate=0]{110};
\draw [anchor=base] (18,0.825) node[scale=1.66573, color=c, rotate=0]{120};
\draw [anchor= east] (18,0.3) node[scale=1.66573, color=c, rotate=0]{$\text{Proton range}\,\,R_\mathrm{w}\,/\,\text{mm}$};
\draw [c,line width=0.9] (2,1.5) -- (2,13.5);
\draw [c,line width=0.9] (2.48,1.5) -- (2,1.5);
\draw [c,line width=0.9] (2.24,1.9) -- (2,1.9);
\draw [c,line width=0.9] (2.24,2.3) -- (2,2.3);
\draw [c,line width=0.9] (2.24,2.7) -- (2,2.7);
\draw [c,line width=0.9] (2.24,3.1) -- (2,3.1);
\draw [c,line width=0.9] (2.48,3.5) -- (2,3.5);
\draw [c,line width=0.9] (2.24,3.9) -- (2,3.9);
\draw [c,line width=0.9] (2.24,4.3) -- (2,4.3);
\draw [c,line width=0.9] (2.24,4.7) -- (2,4.7);
\draw [c,line width=0.9] (2.24,5.1) -- (2,5.1);
\draw [c,line width=0.9] (2.48,5.5) -- (2,5.5);
\draw [c,line width=0.9] (2.24,5.9) -- (2,5.9);
\draw [c,line width=0.9] (2.24,6.3) -- (2,6.3);
\draw [c,line width=0.9] (2.24,6.7) -- (2,6.7);
\draw [c,line width=0.9] (2.24,7.1) -- (2,7.1);
\draw [c,line width=0.9] (2.48,7.5) -- (2,7.5);
\draw [c,line width=0.9] (2.24,7.9) -- (2,7.9);
\draw [c,line width=0.9] (2.24,8.3) -- (2,8.3);
\draw [c,line width=0.9] (2.24,8.7) -- (2,8.7);
\draw [c,line width=0.9] (2.24,9.1) -- (2,9.1);
\draw [c,line width=0.9] (2.48,9.5) -- (2,9.5);
\draw [c,line width=0.9] (2.24,9.9) -- (2,9.9);
\draw [c,line width=0.9] (2.24,10.3) -- (2,10.3);
\draw [c,line width=0.9] (2.24,10.7) -- (2,10.7);
\draw [c,line width=0.9] (2.24,11.1) -- (2,11.1);
\draw [c,line width=0.9] (2.48,11.5) -- (2,11.5);
\draw [c,line width=0.9] (2.24,11.9) -- (2,11.9);
\draw [c,line width=0.9] (2.24,12.3) -- (2,12.3);
\draw [c,line width=0.9] (2.24,12.7) -- (2,12.7);
\draw [c,line width=0.9] (2.24,13.1) -- (2,13.1);
\draw [c,line width=0.9] (2.48,13.5) -- (2,13.5);
\draw [anchor= east] (1.9,1.5) node[scale=1.66573, color=c, rotate=0]{0};
\draw [anchor= east] (1.9,3.5) node[scale=1.66573, color=c, rotate=0]{1};
\draw [anchor= east] (1.9,5.5) node[scale=1.66573, color=c, rotate=0]{2};
\draw [anchor= east] (1.9,7.5) node[scale=1.66573, color=c, rotate=0]{3};
\draw [anchor= east] (1.9,9.5) node[scale=1.66573, color=c, rotate=0]{4};
\draw [anchor= east] (1.9,11.5) node[scale=1.66573, color=c, rotate=0]{5};
\draw [anchor= east] (1.9,13.5) node[scale=1.66573, color=c, rotate=0]{6};
\draw [anchor= east] (0.633333,13.5) node[scale=1.66573, color=c, rotate=90]{$\text{Relative emission variation}\,\,\Delta_{\gamma}(\delta_{R}=1\,\text{mm})\,/\,\,\\\%$};
\draw [c,line width=1.8] (2.8,9.49992) -- (2.96,9.19223) -- (3.12,8.90734) -- (3.28,8.64279) -- (3.44,8.39649) -- (3.6,8.1666) -- (3.76,7.95155) -- (3.92,7.74994) -- (4.08,7.56055) -- (4.24,7.3823) -- (4.4,7.21423) -- (4.56,7.0555) -- (4.72,6.90535)
 -- (4.88,6.76311) -- (5.04,6.62816) -- (5.2,6.49995) -- (5.36,6.378) -- (5.52,6.26186) -- (5.68,6.15112) -- (5.84,6.04541) -- (6,5.9444) -- (6.16,5.84778) -- (6.32,5.75528) -- (6.48,5.66663) -- (6.64,5.58159) -- (6.8,5.49996) -- (6.96,5.42153) --
 (7.12,5.34612) -- (7.28,5.27355) -- (7.44,5.20367) -- (7.6,5.13633) -- (7.76,5.07139) -- (7.92,5.00874) -- (8.08,4.94824) -- (8.24,4.8898) -- (8.4,4.8333) -- (8.56,4.77866) -- (8.72,4.72578) -- (8.88,4.67457) -- (9.04,4.62497) -- (9.2,4.57689) --
 (9.36,4.53027) -- (9.52,4.48505) -- (9.68,4.44115) -- (9.84,4.39852) -- (10,4.35712) -- (10.16,4.31687) -- (10.32,4.27775) -- (10.48,4.2397) -- (10.64,4.20268) -- (10.8,4.16664) -- (10.96,4.13155) -- (11.12,4.09738) -- (11.28,4.06408) --
 (11.44,4.03162) -- (11.6,3.99998) -- (11.76,3.96911) -- (11.92,3.939) -- (12.08,3.90962) -- (12.24,3.88093) -- (12.4,3.85292) -- (12.56,3.82556) -- (12.72,3.79883) -- (12.88,3.77271) -- (13.04,3.74717) -- (13.2,3.7222) -- (13.36,3.69778) --
 (13.52,3.67389) -- (13.68,3.65052) -- (13.84,3.62764) -- (14,3.60524) -- (14.16,3.58331) -- (14.32,3.56184) -- (14.48,3.5408) -- (14.64,3.52018) -- (14.8,3.49998) -- (14.96,3.48018) -- (15.12,3.46077) -- (15.28,3.44173) -- (15.44,3.42306) --
 (15.6,3.40474) -- (15.76,3.38677) -- (15.92,3.36914) -- (16.08,3.35183) -- (16.24,3.33484) -- (16.4,3.31816) -- (16.56,3.30178) -- (16.72,3.2857) -- (16.88,3.26989) -- (17.04,3.25437) -- (17.2,3.23911);
\draw [c,dash pattern=on 2.40pt off 2.40pt ,line width=0.9] (2.8,6.7116) -- (2.96,6.59049) -- (3.12,6.47571) -- (3.28,6.3667) -- (3.44,6.26295) -- (3.6,6.16403) -- (3.76,6.06954) -- (3.92,5.97915) -- (4.08,5.89253) -- (4.24,5.80941) -- (4.4,5.72955)
 -- (4.56,5.6527) -- (4.72,5.57868) -- (4.88,5.50728) -- (5.04,5.43835) -- (5.2,5.37173) -- (5.36,5.30727) -- (5.52,5.24485) -- (5.68,5.18434) -- (5.84,5.12565) -- (6,5.06866) -- (6.16,5.01328) -- (6.32,4.95942) -- (6.48,4.90701) -- (6.64,4.85597) --
 (6.8,4.80623) -- (6.96,4.75773) -- (7.12,4.7104) -- (7.28,4.66418) -- (7.44,4.61903) -- (7.6,4.5749) -- (7.76,4.53174) -- (7.92,4.48949) -- (8.08,4.44814) -- (8.24,4.40762) -- (8.4,4.36791) -- (8.56,4.32898) -- (8.72,4.29078) -- (8.88,4.2533) --
 (9.04,4.21649) -- (9.2,4.18033) -- (9.36,4.1448) -- (9.52,4.10987) -- (9.68,4.07551) -- (9.84,4.04171) -- (10,4.00843) -- (10.16,3.97567) -- (10.32,3.9434) -- (10.48,3.9116) -- (10.64,3.88025) -- (10.8,3.84934) -- (10.96,3.81885) -- (11.12,3.78877)
 -- (11.28,3.75907) -- (11.44,3.72974) -- (11.6,3.70078) -- (11.76,3.67216) -- (11.92,3.64387) -- (12.08,3.6159) -- (12.24,3.58824) -- (12.4,3.56088) -- (12.56,3.5338) -- (12.72,3.50699) -- (12.88,3.48045) -- (13.04,3.45415) -- (13.2,3.4281) --
 (13.36,3.40228) -- (13.52,3.37669) -- (13.68,3.3513) -- (13.84,3.32612) -- (14,3.30114) -- (14.16,3.27634) -- (14.32,3.25171) -- (14.48,3.22726) -- (14.64,3.20296) -- (14.8,3.17882) -- (14.96,3.15482) -- (15.12,3.13096) -- (15.28,3.10722) --
 (15.44,3.08361) -- (15.6,3.06011) -- (15.76,3.03671) -- (15.92,3.01341) -- (16.08,2.9902) -- (16.24,2.96706) -- (16.4,2.94401) -- (16.56,2.92102) -- (16.72,2.89809) -- (16.88,2.87521) -- (17.04,2.85237) -- (17.2,2.82957);
\end{tikzpicture}
 }
\vspace{-12pt}
\caption{Solid line: Relative change $\Delta_\gamma$ of emitted \grs as a function of proton range $R_\mathrm{w}$ for an error $\delta_R=\text{1}\,\text{mm}$, \cf \refeq{eq:deltaG}. Dashed line: Statistical measurement precision $\Delta_\mathrm{s}$, \cf \refeq{eq:deltaS}, for the coaxial detection setup described in \refig{fig:setup} and table~\ref{tab:params}, and for a spot cluster with $N_\mathrm{p}=\text{10}^\text{8}$ protons.}
\label{fig:gammaDeltaG}

\vspace{8pt}
\hspace{6mm}\scalebox{0.4}{\begin{tikzpicture}
\pgfdeclareplotmark{cross} {
\pgfpathmoveto{\pgfpoint{-0.3\pgfplotmarksize}{\pgfplotmarksize}}
\pgfpathlineto{\pgfpoint{+0.3\pgfplotmarksize}{\pgfplotmarksize}}
\pgfpathlineto{\pgfpoint{+0.3\pgfplotmarksize}{0.3\pgfplotmarksize}}
\pgfpathlineto{\pgfpoint{+1\pgfplotmarksize}{0.3\pgfplotmarksize}}
\pgfpathlineto{\pgfpoint{+1\pgfplotmarksize}{-0.3\pgfplotmarksize}}
\pgfpathlineto{\pgfpoint{+0.3\pgfplotmarksize}{-0.3\pgfplotmarksize}}
\pgfpathlineto{\pgfpoint{+0.3\pgfplotmarksize}{-1.\pgfplotmarksize}}
\pgfpathlineto{\pgfpoint{-0.3\pgfplotmarksize}{-1.\pgfplotmarksize}}
\pgfpathlineto{\pgfpoint{-0.3\pgfplotmarksize}{-0.3\pgfplotmarksize}}
\pgfpathlineto{\pgfpoint{-1.\pgfplotmarksize}{-0.3\pgfplotmarksize}}
\pgfpathlineto{\pgfpoint{-1.\pgfplotmarksize}{0.3\pgfplotmarksize}}
\pgfpathlineto{\pgfpoint{-0.3\pgfplotmarksize}{0.3\pgfplotmarksize}}
\pgfpathclose
\pgfusepathqstroke
}
\pgfdeclareplotmark{cross*} {
\pgfpathmoveto{\pgfpoint{-0.3\pgfplotmarksize}{\pgfplotmarksize}}
\pgfpathlineto{\pgfpoint{+0.3\pgfplotmarksize}{\pgfplotmarksize}}
\pgfpathlineto{\pgfpoint{+0.3\pgfplotmarksize}{0.3\pgfplotmarksize}}
\pgfpathlineto{\pgfpoint{+1\pgfplotmarksize}{0.3\pgfplotmarksize}}
\pgfpathlineto{\pgfpoint{+1\pgfplotmarksize}{-0.3\pgfplotmarksize}}
\pgfpathlineto{\pgfpoint{+0.3\pgfplotmarksize}{-0.3\pgfplotmarksize}}
\pgfpathlineto{\pgfpoint{+0.3\pgfplotmarksize}{-1.\pgfplotmarksize}}
\pgfpathlineto{\pgfpoint{-0.3\pgfplotmarksize}{-1.\pgfplotmarksize}}
\pgfpathlineto{\pgfpoint{-0.3\pgfplotmarksize}{-0.3\pgfplotmarksize}}
\pgfpathlineto{\pgfpoint{-1.\pgfplotmarksize}{-0.3\pgfplotmarksize}}
\pgfpathlineto{\pgfpoint{-1.\pgfplotmarksize}{0.3\pgfplotmarksize}}
\pgfpathlineto{\pgfpoint{-0.3\pgfplotmarksize}{0.3\pgfplotmarksize}}
\pgfpathclose
\pgfusepathqfillstroke
}
\pgfdeclareplotmark{newstar} {
\pgfpathmoveto{\pgfqpoint{0pt}{\pgfplotmarksize}}
\pgfpathlineto{\pgfqpointpolar{44}{0.5\pgfplotmarksize}}
\pgfpathlineto{\pgfqpointpolar{18}{\pgfplotmarksize}}
\pgfpathlineto{\pgfqpointpolar{-20}{0.5\pgfplotmarksize}}
\pgfpathlineto{\pgfqpointpolar{-54}{\pgfplotmarksize}}
\pgfpathlineto{\pgfqpointpolar{-90}{0.5\pgfplotmarksize}}
\pgfpathlineto{\pgfqpointpolar{234}{\pgfplotmarksize}}
\pgfpathlineto{\pgfqpointpolar{198}{0.5\pgfplotmarksize}}
\pgfpathlineto{\pgfqpointpolar{162}{\pgfplotmarksize}}
\pgfpathlineto{\pgfqpointpolar{134}{0.5\pgfplotmarksize}}
\pgfpathclose
\pgfusepathqstroke
}
\pgfdeclareplotmark{newstar*} {
\pgfpathmoveto{\pgfqpoint{0pt}{\pgfplotmarksize}}
\pgfpathlineto{\pgfqpointpolar{44}{0.5\pgfplotmarksize}}
\pgfpathlineto{\pgfqpointpolar{18}{\pgfplotmarksize}}
\pgfpathlineto{\pgfqpointpolar{-20}{0.5\pgfplotmarksize}}
\pgfpathlineto{\pgfqpointpolar{-54}{\pgfplotmarksize}}
\pgfpathlineto{\pgfqpointpolar{-90}{0.5\pgfplotmarksize}}
\pgfpathlineto{\pgfqpointpolar{234}{\pgfplotmarksize}}
\pgfpathlineto{\pgfqpointpolar{198}{0.5\pgfplotmarksize}}
\pgfpathlineto{\pgfqpointpolar{162}{\pgfplotmarksize}}
\pgfpathlineto{\pgfqpointpolar{134}{0.5\pgfplotmarksize}}
\pgfpathclose
\pgfusepathqfillstroke
}
\definecolor{c}{rgb}{0,0,0};
\draw [c,line width=0.9] (2,1.5) -- (2,13.5) -- (18,13.5) -- (18,1.5) -- (2,1.5);
\draw [c,line width=0.9] (2,1.5) -- (2,13.5) -- (18,13.5) -- (18,1.5) -- (2,1.5);
\draw [c,line width=0.9] (2,1.5) -- (2,13.5) -- (18,13.5) -- (18,1.5) -- (2,1.5);
\draw [c,line width=0.9] (2,1.5) -- (2,13.5) -- (18,13.5) -- (18,1.5) -- (2,1.5);
\draw [c,line width=0.9] (2,1.5) -- (18,1.5);
\draw [c,dash pattern=on 0.80pt off 1.60pt ,line width=0.9] (2,13.5) -- (2,1.5);
\draw [c,dash pattern=on 0.80pt off 1.60pt ,line width=0.9] (3.6,13.5) -- (3.6,1.5);
\draw [c,dash pattern=on 0.80pt off 1.60pt ,line width=0.9] (5.2,13.5) -- (5.2,1.5);
\draw [c,dash pattern=on 0.80pt off 1.60pt ,line width=0.9] (6.8,13.5) -- (6.8,1.5);
\draw [c,dash pattern=on 0.80pt off 1.60pt ,line width=0.9] (8.4,13.5) -- (8.4,1.5);
\draw [c,dash pattern=on 0.80pt off 1.60pt ,line width=0.9] (10,13.5) -- (10,1.5);
\draw [c,dash pattern=on 0.80pt off 1.60pt ,line width=0.9] (11.6,13.5) -- (11.6,1.5);
\draw [c,dash pattern=on 0.80pt off 1.60pt ,line width=0.9] (13.2,13.5) -- (13.2,1.5);
\draw [c,dash pattern=on 0.80pt off 1.60pt ,line width=0.9] (14.8,13.5) -- (14.8,1.5);
\draw [c,dash pattern=on 0.80pt off 1.60pt ,line width=0.9] (16.4,13.5) -- (16.4,1.5);
\draw [c,dash pattern=on 0.80pt off 1.60pt ,line width=0.9] (18,13.5) -- (18,1.5);
\draw [c,line width=0.9] (2,1.5) -- (2,13.5);
\draw [c,dash pattern=on 0.80pt off 1.60pt ,line width=0.9] (18,1.5) -- (2,1.5);
\draw [c,dash pattern=on 0.80pt off 1.60pt ,line width=0.9] (18,3.5) -- (2,3.5);
\draw [c,dash pattern=on 0.80pt off 1.60pt ,line width=0.9] (18,5.5) -- (2,5.5);
\draw [c,dash pattern=on 0.80pt off 1.60pt ,line width=0.9] (18,7.5) -- (2,7.5);
\draw [c,dash pattern=on 0.80pt off 1.60pt ,line width=0.9] (18,9.5) -- (2,9.5);
\draw [c,dash pattern=on 0.80pt off 1.60pt ,line width=0.9] (18,11.5) -- (2,11.5);
\draw [c,dash pattern=on 0.80pt off 1.60pt ,line width=0.9] (18,13.5) -- (2,13.5);
\draw [c,line width=0.9] (2,1.5) -- (18,1.5);
\draw [c,line width=0.9] (2,1.86) -- (2,1.5);
\draw [c,line width=0.9] (2.32,1.68) -- (2.32,1.5);
\draw [c,line width=0.9] (2.64,1.68) -- (2.64,1.5);
\draw [c,line width=0.9] (2.96,1.68) -- (2.96,1.5);
\draw [c,line width=0.9] (3.28,1.68) -- (3.28,1.5);
\draw [c,line width=0.9] (3.6,1.86) -- (3.6,1.5);
\draw [c,line width=0.9] (3.92,1.68) -- (3.92,1.5);
\draw [c,line width=0.9] (4.24,1.68) -- (4.24,1.5);
\draw [c,line width=0.9] (4.56,1.68) -- (4.56,1.5);
\draw [c,line width=0.9] (4.88,1.68) -- (4.88,1.5);
\draw [c,line width=0.9] (5.2,1.86) -- (5.2,1.5);
\draw [c,line width=0.9] (5.52,1.68) -- (5.52,1.5);
\draw [c,line width=0.9] (5.84,1.68) -- (5.84,1.5);
\draw [c,line width=0.9] (6.16,1.68) -- (6.16,1.5);
\draw [c,line width=0.9] (6.48,1.68) -- (6.48,1.5);
\draw [c,line width=0.9] (6.8,1.86) -- (6.8,1.5);
\draw [c,line width=0.9] (7.12,1.68) -- (7.12,1.5);
\draw [c,line width=0.9] (7.44,1.68) -- (7.44,1.5);
\draw [c,line width=0.9] (7.76,1.68) -- (7.76,1.5);
\draw [c,line width=0.9] (8.08,1.68) -- (8.08,1.5);
\draw [c,line width=0.9] (8.4,1.86) -- (8.4,1.5);
\draw [c,line width=0.9] (8.72,1.68) -- (8.72,1.5);
\draw [c,line width=0.9] (9.04,1.68) -- (9.04,1.5);
\draw [c,line width=0.9] (9.36,1.68) -- (9.36,1.5);
\draw [c,line width=0.9] (9.68,1.68) -- (9.68,1.5);
\draw [c,line width=0.9] (10,1.86) -- (10,1.5);
\draw [c,line width=0.9] (10.32,1.68) -- (10.32,1.5);
\draw [c,line width=0.9] (10.64,1.68) -- (10.64,1.5);
\draw [c,line width=0.9] (10.96,1.68) -- (10.96,1.5);
\draw [c,line width=0.9] (11.28,1.68) -- (11.28,1.5);
\draw [c,line width=0.9] (11.6,1.86) -- (11.6,1.5);
\draw [c,line width=0.9] (11.92,1.68) -- (11.92,1.5);
\draw [c,line width=0.9] (12.24,1.68) -- (12.24,1.5);
\draw [c,line width=0.9] (12.56,1.68) -- (12.56,1.5);
\draw [c,line width=0.9] (12.88,1.68) -- (12.88,1.5);
\draw [c,line width=0.9] (13.2,1.86) -- (13.2,1.5);
\draw [c,line width=0.9] (13.52,1.68) -- (13.52,1.5);
\draw [c,line width=0.9] (13.84,1.68) -- (13.84,1.5);
\draw [c,line width=0.9] (14.16,1.68) -- (14.16,1.5);
\draw [c,line width=0.9] (14.48,1.68) -- (14.48,1.5);
\draw [c,line width=0.9] (14.8,1.86) -- (14.8,1.5);
\draw [c,line width=0.9] (15.12,1.68) -- (15.12,1.5);
\draw [c,line width=0.9] (15.44,1.68) -- (15.44,1.5);
\draw [c,line width=0.9] (15.76,1.68) -- (15.76,1.5);
\draw [c,line width=0.9] (16.08,1.68) -- (16.08,1.5);
\draw [c,line width=0.9] (16.4,1.86) -- (16.4,1.5);
\draw [c,line width=0.9] (16.72,1.68) -- (16.72,1.5);
\draw [c,line width=0.9] (17.04,1.68) -- (17.04,1.5);
\draw [c,line width=0.9] (17.36,1.68) -- (17.36,1.5);
\draw [c,line width=0.9] (17.68,1.68) -- (17.68,1.5);
\draw [c,line width=0.9] (18,1.86) -- (18,1.5);
\draw [anchor=base] (2,0.825) node[scale=1.66573, color=c, rotate=0]{20};
\draw [anchor=base] (3.6,0.825) node[scale=1.66573, color=c, rotate=0]{30};
\draw [anchor=base] (5.2,0.825) node[scale=1.66573, color=c, rotate=0]{40};
\draw [anchor=base] (6.8,0.825) node[scale=1.66573, color=c, rotate=0]{50};
\draw [anchor=base] (8.4,0.825) node[scale=1.66573, color=c, rotate=0]{60};
\draw [anchor=base] (10,0.825) node[scale=1.66573, color=c, rotate=0]{70};
\draw [anchor=base] (11.6,0.825) node[scale=1.66573, color=c, rotate=0]{80};
\draw [anchor=base] (13.2,0.825) node[scale=1.66573, color=c, rotate=0]{90};
\draw [anchor=base] (14.8,0.825) node[scale=1.66573, color=c, rotate=0]{100};
\draw [anchor=base] (16.4,0.825) node[scale=1.66573, color=c, rotate=0]{110};
\draw [anchor=base] (18,0.825) node[scale=1.66573, color=c, rotate=0]{120};
\draw [anchor= east] (18,0.3) node[scale=1.66573, color=c, rotate=0]{$\text{Proton range}\,\,R_\mathrm{w}\,/\,\text{mm}$};
\draw [c,line width=0.9] (2,1.5) -- (2,13.5);
\draw [c,line width=0.9] (2.48,1.5) -- (2,1.5);
\draw [c,line width=0.9] (2.24,1.9) -- (2,1.9);
\draw [c,line width=0.9] (2.24,2.3) -- (2,2.3);
\draw [c,line width=0.9] (2.24,2.7) -- (2,2.7);
\draw [c,line width=0.9] (2.24,3.1) -- (2,3.1);
\draw [c,line width=0.9] (2.48,3.5) -- (2,3.5);
\draw [c,line width=0.9] (2.24,3.9) -- (2,3.9);
\draw [c,line width=0.9] (2.24,4.3) -- (2,4.3);
\draw [c,line width=0.9] (2.24,4.7) -- (2,4.7);
\draw [c,line width=0.9] (2.24,5.1) -- (2,5.1);
\draw [c,line width=0.9] (2.48,5.5) -- (2,5.5);
\draw [c,line width=0.9] (2.24,5.9) -- (2,5.9);
\draw [c,line width=0.9] (2.24,6.3) -- (2,6.3);
\draw [c,line width=0.9] (2.24,6.7) -- (2,6.7);
\draw [c,line width=0.9] (2.24,7.1) -- (2,7.1);
\draw [c,line width=0.9] (2.48,7.5) -- (2,7.5);
\draw [c,line width=0.9] (2.24,7.9) -- (2,7.9);
\draw [c,line width=0.9] (2.24,8.3) -- (2,8.3);
\draw [c,line width=0.9] (2.24,8.7) -- (2,8.7);
\draw [c,line width=0.9] (2.24,9.1) -- (2,9.1);
\draw [c,line width=0.9] (2.48,9.5) -- (2,9.5);
\draw [c,line width=0.9] (2.24,9.9) -- (2,9.9);
\draw [c,line width=0.9] (2.24,10.3) -- (2,10.3);
\draw [c,line width=0.9] (2.24,10.7) -- (2,10.7);
\draw [c,line width=0.9] (2.24,11.1) -- (2,11.1);
\draw [c,line width=0.9] (2.48,11.5) -- (2,11.5);
\draw [c,line width=0.9] (2.24,11.9) -- (2,11.9);
\draw [c,line width=0.9] (2.24,12.3) -- (2,12.3);
\draw [c,line width=0.9] (2.24,12.7) -- (2,12.7);
\draw [c,line width=0.9] (2.24,13.1) -- (2,13.1);
\draw [c,line width=0.9] (2.48,13.5) -- (2,13.5);
\draw [anchor= east] (1.9,1.5) node[scale=1.66573, color=c, rotate=0]{0};
\draw [anchor= east] (1.9,3.5) node[scale=1.66573, color=c, rotate=0]{1};
\draw [anchor= east] (1.9,5.5) node[scale=1.66573, color=c, rotate=0]{2};
\draw [anchor= east] (1.9,7.5) node[scale=1.66573, color=c, rotate=0]{3};
\draw [anchor= east] (1.9,9.5) node[scale=1.66573, color=c, rotate=0]{4};
\draw [anchor= east] (1.9,11.5) node[scale=1.66573, color=c, rotate=0]{5};
\draw [anchor= east] (1.9,13.5) node[scale=1.66573, color=c, rotate=0]{6};
\draw [anchor= east] (0.633333,13.5) node[scale=1.66573, color=c, rotate=90]{$\text{Relative count variation}\,\,\Delta_{\Omega}(\delta_{z}=1\,\text{mm})\,/\,\,\\\%$};
\draw [c,line width=1.8] (2.8,4.37602) -- (2.96,4.38822) -- (3.12,4.40062) -- (3.28,4.41322) -- (3.44,4.42603) -- (3.6,4.43905) -- (3.76,4.45229) -- (3.92,4.46575) -- (4.08,4.47943) -- (4.24,4.49335) -- (4.4,4.5075) -- (4.56,4.5219) -- (4.72,4.53654)
 -- (4.88,4.55144) -- (5.04,4.5666) -- (5.2,4.58203) -- (5.36,4.59774) -- (5.52,4.61372) -- (5.68,4.63) -- (5.84,4.64657) -- (6,4.66345) -- (6.16,4.68064) -- (6.32,4.69815) -- (6.48,4.716) -- (6.64,4.73418) -- (6.8,4.75271) -- (6.96,4.77161) --
 (7.12,4.79087) -- (7.28,4.81052) -- (7.44,4.83056) -- (7.6,4.851) -- (7.76,4.87186) -- (7.92,4.89315) -- (8.08,4.91488) -- (8.24,4.93706) -- (8.4,4.95972) -- (8.56,4.98286) -- (8.72,5.0065) -- (8.88,5.03066) -- (9.04,5.05535) -- (9.2,5.0806) --
 (9.36,5.10641) -- (9.52,5.13281) -- (9.68,5.15982) -- (9.84,5.18747) -- (10,5.21576) -- (10.16,5.24473) -- (10.32,5.2744) -- (10.48,5.3048) -- (10.64,5.33594) -- (10.8,5.36787) -- (10.96,5.4006) -- (11.12,5.43417) -- (11.28,5.46861) --
 (11.44,5.50396) -- (11.6,5.54025) -- (11.76,5.57752) -- (11.92,5.6158) -- (12.08,5.65514) -- (12.24,5.69559) -- (12.4,5.73718) -- (12.56,5.77998) -- (12.72,5.82402) -- (12.88,5.86936) -- (13.04,5.91607) -- (13.2,5.96419) -- (13.36,6.01381) --
 (13.52,6.06497) -- (13.68,6.11777) -- (13.84,6.17227) -- (14,6.22856) -- (14.16,6.28671) -- (14.32,6.34684) -- (14.48,6.40902) -- (14.64,6.47338) -- (14.8,6.54002) -- (14.96,6.60905) -- (15.12,6.68062) -- (15.28,6.75484) -- (15.44,6.83188) --
 (15.6,6.91189) -- (15.76,6.99502) -- (15.92,7.08148) -- (16.08,7.17143) -- (16.24,7.2651) -- (16.4,7.36271) -- (16.56,7.4645) -- (16.72,7.57072) -- (16.88,7.68165) -- (17.04,7.79759) -- (17.2,7.91888);
\draw [c,dash pattern=on 2.40pt off 2.40pt ,line width=0.9] (2.8,6.7116) -- (2.96,6.59049) -- (3.12,6.47571) -- (3.28,6.3667) -- (3.44,6.26295) -- (3.6,6.16403) -- (3.76,6.06954) -- (3.92,5.97915) -- (4.08,5.89253) -- (4.24,5.80941) -- (4.4,5.72955)
 -- (4.56,5.6527) -- (4.72,5.57868) -- (4.88,5.50728) -- (5.04,5.43835) -- (5.2,5.37173) -- (5.36,5.30727) -- (5.52,5.24485) -- (5.68,5.18434) -- (5.84,5.12565) -- (6,5.06866) -- (6.16,5.01328) -- (6.32,4.95942) -- (6.48,4.90701) -- (6.64,4.85597) --
 (6.8,4.80623) -- (6.96,4.75773) -- (7.12,4.7104) -- (7.28,4.66418) -- (7.44,4.61903) -- (7.6,4.5749) -- (7.76,4.53174) -- (7.92,4.48949) -- (8.08,4.44814) -- (8.24,4.40762) -- (8.4,4.36791) -- (8.56,4.32898) -- (8.72,4.29078) -- (8.88,4.2533) --
 (9.04,4.21649) -- (9.2,4.18033) -- (9.36,4.1448) -- (9.52,4.10987) -- (9.68,4.07551) -- (9.84,4.04171) -- (10,4.00843) -- (10.16,3.97567) -- (10.32,3.9434) -- (10.48,3.9116) -- (10.64,3.88025) -- (10.8,3.84934) -- (10.96,3.81885) -- (11.12,3.78877)
 -- (11.28,3.75907) -- (11.44,3.72974) -- (11.6,3.70078) -- (11.76,3.67216) -- (11.92,3.64387) -- (12.08,3.6159) -- (12.24,3.58824) -- (12.4,3.56088) -- (12.56,3.5338) -- (12.72,3.50699) -- (12.88,3.48045) -- (13.04,3.45415) -- (13.2,3.4281) --
 (13.36,3.40228) -- (13.52,3.37669) -- (13.68,3.3513) -- (13.84,3.32612) -- (14,3.30114) -- (14.16,3.27634) -- (14.32,3.25171) -- (14.48,3.22726) -- (14.64,3.20296) -- (14.8,3.17882) -- (14.96,3.15482) -- (15.12,3.13096) -- (15.28,3.10722) --
 (15.44,3.08361) -- (15.6,3.06011) -- (15.76,3.03671) -- (15.92,3.01341) -- (16.08,2.9902) -- (16.24,2.96706) -- (16.4,2.94401) -- (16.56,2.92102) -- (16.72,2.89809) -- (16.88,2.87521) -- (17.04,2.85237) -- (17.2,2.82957);
\end{tikzpicture}
 }
\vspace{-12pt}
\caption{Solid line: Relative change $\Delta_\Omega$ of measured \grs as a function of proton range $R_\mathrm{w}$ for an error $\delta_z=\text{1}\,\text{mm}$ in phantom position, \cf \refeq{eq:deltaO}, for the coaxial detection setup described in \refig{fig:setup} and table~\ref{tab:params}. Dashed line: Statistical measurement precision $\Delta_\mathrm{s}$, \cf \refeq{eq:deltaS}, for a spot cluster with $N_\mathrm{p}=\text{10}^\text{8}$ protons.}
\label{fig:gammaDeltaO}
\end{figure}

\begin{figure}[h]
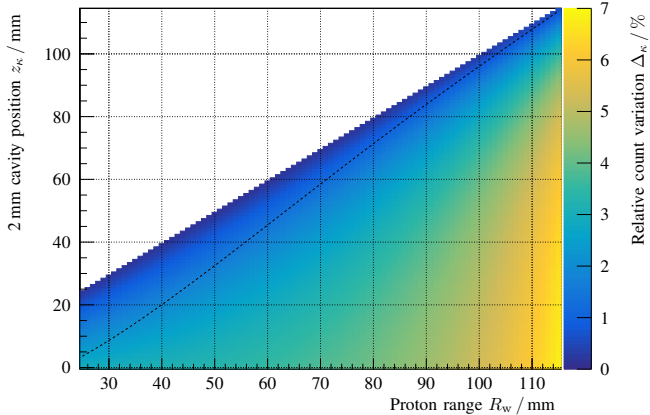

\ifthenelse{ \equal{\rapido}{true} }{}{

\scalebox{0.4}{

 }}
\caption{Relative change $\Delta_\kappa$ of measured \grs after introduction of a 2\,mm cavity in the beam path, as a function of original proton range $R_\mathrm{w}$ and cavity position $z_\kappa$, \cf \refeq{eq:deltaK}, for the coaxial detection setup described in \refig{fig:setup} and table~\ref{tab:params}. The black dashed line represents the statistical measurement precision $\Delta_\mathrm{s}$, \cf \refeq{eq:deltaS}, for a spot cluster with $N_\mathrm{p}=\text{10}^\text{8}$ protons.}
\label{fig:gammaCavities}
\end{figure}

\newpage
The variation in detected counts $\Delta_\kappa$ due to the introduction of an air cavity (thickness $\kappa=\text{2}$\,mm) in beam direction with respect to the homogeneous scenario is presented in \refig{fig:gammaCavities}. The dashed black curve represents the boundary above which the expected count variation $\Delta_\kappa$ is below the statistical measurement precision $\Delta_\mathrm{s}$ at that abscissa $R_\mathrm{w}$. For the majority of proton ranges $R_\mathrm{w}$ and cavity positions $z_\kappa$, the expected count change is higher than the statistical measurement uncertainty for $N_\mathrm{p}=\text{10}^\text{8}$ delivered protons.

In \refig{fig:gammaProduction}, the number of prompt \grs emitted per proton as obtained with a TOPAS \MC simulation (black dots) is depicted and compared with the proportional estimation $\bar{g}_\lambda R_\mathrm{w}$ of the analytical model (red solid line), \cf \refeq{eq:gammaint}. The model shape, with a fixed parameter $\bar{g}_\lambda=6\times10^{-4}$, is a rough first-order approximation of the simulated depth-emission profile obtained. The model underestimates the slope of the curve by 30\,\% for small proton ranges and overestimates it by 30\,\% at larger proton ranges.

The choice of a constant value $\bar{g}_\lambda$ implies that the predicted prompt \gr emission distribution is flat, except for the smooth falloff near the particle range according to parameter $\bar{f}$, \cf \refeq{eq:dNe}. For an incident proton energy of 100\,MeV, this approximation (solid red line) is compared with a TOPAS \MC simulation (black dots) of the differential prompt \gr production, \cf \refig{fig:BraggZ}. The modeled profile overestimates the emission about 20\,\%  from the TOPAS simulation and underestimates it by 30\,\% close to the proton range $R_\mathrm{w}$ (vertical purple line).

\begin{figure}[t]
\hspace{5mm}\scalebox{0.4}{

 }
\caption{Total number of prompt \grs $N_\gamma$ (all energies) emitted per proton as a function of the beam range $R_\mathrm{w}$ in the water phantom described in \refig{fig:setup} and table~\ref{tab:params}. The corresponding beam energy is indicated in the upper horizontal axis. The points with error bars represent the results obtained with a TOPAS simulation and $N_\mathrm{p}=\,$10\tup{8} histories, while the solid red line indicates the analytical model prediction, \cf \refeq{eq:gammaint} and table~\ref{tab:params}.}
\label{fig:gammaProduction}

\vspace{5mm}
\hspace{6mm}\scalebox{0.4}{

 }
\caption{Number of prompt \grs $N_\gamma$ (all energies) emitted per proton and unit length as a function of the penetration depth $z-z_0$ within the water phantom described in \refig{fig:setup} and table~\ref{tab:params}. The incident beam energy is 100\,MeV, corresponding to a proton range in water of $R_\mathrm{w}=$\,76.4\,mm (vertical purple line). The points with error bars represent the results obtained with a TOPAS simulation and $N_\mathrm{p}=\,$10\tup{8} histories, while the solid red line indicates the analytical model prediction, \cf \refeq{eq:dNe} and table~\ref{tab:params}.}
\label{fig:BraggZ}
\end{figure}

The ratio between all events detected and the subset of those that are prompt \grs is shown in \refig{fig:gammaEvRatio}, where the TOPAS \MC simulation (black dots) is compared with the constant estimate $k_\gamma=2.5$ of the analytical model (solid red line), \cf subsection~\ref{ss:background}. The model overestimates the amount (ratio) of background for all proton ranges; the smaller the proton range, the higher the deviation with respect to the simulation.
A similar trend is observed in \refig{fig:gammaDet}, where the detector count rates due to background (green) and prompt \grs (blue) are shown for both the \MC simulation and the analytical model, \cf \refeq{eq:dload}.

\begin{figure}[t]
\hspace{5mm}\scalebox{0.4}{

 }
\caption{Ratio $k_\gamma$ between all interactions and the subset of those corresponding to prompt \grs for the coaxial detection setup described in \refig{fig:setup} and table~\ref{tab:params}. The points with error bars represent the results obtained with a TOPAS simulation of $N_\mathrm{p}=\,$10\tup{8} histories and an energy deposit threshold of 50\,keV, while the solid red line indicates the prediction according to the analytical \ac{1D} model, \cf subsection~\ref{ss:background}.}
\label{fig:gammaEvRatio}

\hspace{5mm}\scalebox{0.4}{%
 }
\caption{Partial detector count rate $\dot{N}_{\mathrm{d,}x}$ due to $x=\,$background (green) or $x=\,$prompt \grs (blue) for the coaxial detection setup described in \refig{fig:setup} and table~\ref{tab:params}. The points with error bars represent the results obtained with a TOPAS simulation of $N_\mathrm{p}=\,$10\tup{8} histories and an energy deposit threshold of 50\,keV, while the solid red line indicates the prediction according to the analytical \ac{1D} model, \cf \refeq{eq:dload}.}
\label{fig:gammaDet}
\end{figure}
 
\section{Discussion}\label{sec:discussion}

The proposed method for range verification in proton therapy exploits the strong dependency of the integral number of measured \grs on proton range. The key innovation of the \CPGM approach with respect to \PGPI \cite{krimmer_pgpi} is the coaxial detector orientation and the positioning behind the treated area. This is a convenient geometry to maximize the detection efficiency for \grs produced near the Bragg peak. If the number of incident protons, the prompt \gr yield, the phantom material and the detector response are well known, the proton range can be inferred by solely counting the number of prompt \grs detected within a broad energy window.

\subsection{Limitations}\label{ss:limitations}
The formulated \ac{1D} model, \cf section~\ref{sec:theory}, aims at a simplified analytical description of the prompt \gr emission and detection based on rough estimates, in order to assess the feasibility of the \CPGM method. A number of approximations and limitations are identified:

\begin{itemize}
\item \ac{3D} information is not taken into account, \eg the lateral spread of the proton pencil beam as a function of penetration depth.
\item The energy spread of the incident proton beam is not modeled.
\item A simplified \gr emission profile is assumed with a shape independent of proton energy, \cf \refeq{eq:dNe}, and an approximate \gr yield is used.
\item Characteristic \gr lines and Doppler broadening are not considered.
\item \gr anisotropy effects are not incorporated \cite{sheldon_anisotropy},    \cite[Fig.~3]{shute_anisotropy}, \cite[Fig.~2]{kiener_anisotropy}, \cite{mattei_anisotropy}.
\item An ideal homogeneous water phantom is chosen as target.
\item The effect of \CT imaging uncertainties on attenuation coefficients and stopping power conversion are not modeled.
\item The dependency of attenuation coefficients on \gr energy is not taken into account.
\item Scattering of prompt \grs before reaching the detector is neglected, which is however significant at low \gr energies (less forward-peaked) and large distances between scattering point and detector.
\item Detection of prompt annihilation photons, stemming from prompt \grs that undergo pair production in the phantom, is not considered.
\item A simple model of the radiation background is adopted.
\item Neutron interactions and activation in the detector \cite{lu_neutron,cazzaniga_neutron} are not included.
\item The detector response (\gr energy deposit) is not included, and the paraxial approximation is used.
\item Second-order corrections needed for \PBS, as spots are not coaxial (in general) with the room isocenter (detector axis), are not incorporated.
\item A potential systematic error in the incident number of protons is not accounted for.
\item The statistical error in the stochastic process of \gr transmission is not considered.
\item Bias due to systematic errors in the background subtraction is not investigated.
\item Sensitivity to range errors is studied only for simplified cases: global beam range errors (proton energy), target shifts (across beam direction) and local 2\,mm air cavities.
\end{itemize}

\subsection{Sensitivity to range errors}
The sensitivity to range errors is studied for a representative brain tumor irradiation from a horizontal beam angle. With a single $\diameter$1''$\times$1\nicefrac{1}{2}'' scintillation detector, changes of over 3\,\% in the prompt \gr count are expected for an absolute proton range error of $\delta_R=\,$1\,mm, \cf \refig{fig:gammaDeltaN}. In the case of 10\tup{8} delivered protons, the statistical measurement precision would be enough to detect that variation with significance (1$\sigma$).

On the other hand, the proposed setup would only be able to detect a target misalignment across the beam axis of $\delta_z=\,$1\,mm for proton ranges larger than 6\,cm, \cf \refig{fig:gammaDeltaO}. For smaller proton ranges, the sensitivity would be rather 2\,mm, (unless a larger spot cluster or a bigger detector were deployed). A 2\,mm precision would be still an improvement compared with the safety margins currently applied \cite{hueso_fio_cc_pgt,taasti_margins}.

A potential improvement might be the use of a second detector in an upstream orientation and slightly off beam axis, close to the beam entrance point, so that the maximum covered solid angle is in the region where the downstream detector is less sensitive. By combining information from both detectors, complementary information about the absolute target position could be deduced.

The effect of anatomy changes in the patient with respect to the planning \ac{CT} scan \cite{lens_anatomy} is investigated by adding air cavities of $\kappa=\text{2}$\,mm thickness within the beam path, \cf \refig{fig:gammaCavities}. If the cavity is located close to the end of range, the statistical measurement precision is not enough to detect this effect, in part because most of the dose has already been deposited in the expected region. For large proton ranges and cavity locations in the beginning of or amid the beam path, count rate variations of 4\,\% are expected, which are higher than the statistical uncertainty for spots with 10\tup{8} protons.

In the clinical context, where extended targets are irradiated with \PBS, the detector is kept fixed at the beam isocenter. This implies that the scanned pencil beam is no longer coaxial to the detector, which reduces the sensitivity of the \CPGM method to range errors. The geometrical distance between emission point and detector front face is instead $\sqrt{(z_\mathrm{d}-z)^2+\Delta x^2}=(z_\mathrm{d}-z)/\cos\alpha$, where $\Delta x$ is the lateral separation with respect to the beam axis and $\alpha$ is the angle between the $z$ axis and the source-to-detector axis. Hence, at first order in the asymptotic series expansion of \refeq{eq:solid}, the count rate variation $\Delta_\Omega$ due to the solid angle effect is corrected by a factor $\cos^3\alpha=[1+\Delta x^2 / (z_\mathrm{d}-z)^2]^{-1.5}$ within the integrand, \cf \refeq{eq:deltaO}. For example, at a proton range $R_\mathrm{w}=\text{100}$\,mm and a lateral separation $\Delta x = \text{50}$\,mm, the angle $\alpha$ ranges between 10 and 40 degrees across the beam track. Compared to the coaxial case, $\Delta_\Omega$ decreases from 2.5\,\%$/$mm to 1.4\,\%$/$mm, still above the statistical precision of 0.8\,\%, \cf \refig{fig:gammaDeltaO}.

In practice, whether the \CPGM method is able to measure 1\,mm range shifts with statistical significance depends on the lateral distance of the pencil beam to the isocenter, and on the number of protons within the spot cluster. A comparable kind of inhomogeneous precision depending on the considered spot is also present in other \PGI systems oriented perpendicularly to the beam: For spots with the same range and number of protons, but laterally further away from the camera, the solid angle and transmission probability are smaller, so that the number of collected counts (and thus the statistical precision) decreases.

\subsection{Validation}

The incorporation of all effects mentioned in subsection~\ref{ss:limitations} exceeds the scope of this paper and should be subject of either more sophisticated analytical algorithms \cite{sterpin_pg} or a full \MC simulation. It should take into account realistic patient geometries, incorporate treatment room elements to reliably simulate neutron-induced background, as well as model accurately the scintillation detector. Consequently, the results presented are a qualitative measure and indicate the potential of the underlying methodology. To assess whether the limitations of the analytical model could invalidate the measurement principle, a TOPAS \MC simulation is conducted with a simplified geometry, \cf subsection~\ref{ss:validation}, mimicking the water target of \refig{fig:setup}.

The results are compared with the model in view of the most critical approximations and estimates included in the model. If these were off by more than a factor of two, it could jeopardize the feasibility of the \CPGM method. Specifically, the linear approximation in the prompt \gr production is tested in \refig{fig:gammaProduction}, the flat profile $\bar{g}_\lambda$ is studied in \refig{fig:BraggZ}, the constant ratio $k_\gamma$ between all events and prompt \grs is presented in \refig{fig:gammaEvRatio}, and the overall detector count rates are shown in \refig{fig:gammaDet}.

The fact that the model overestimates the slope in the prompt \gr production by 30\,\% at large proton ranges implies a noticeable reduction of the expected count variation $\Delta_\gamma$. However, this is a minor effect at large proton ranges compared to the solid angle contribution $\Delta_\Omega$ in case of a global range error, \cf \refig{fig:gammaDeltaG}. On the other hand, the assumption of a constant ratio $k_\gamma=\,$2.5 is conservative compared to the results obtained with a TOPAS \MC simulation. For all proton ranges, the amount of background obtained with TOPAS \MC is smaller than predicted. Correspondingly, the modeled detector count rates for prompt \grs and background are discrepant with the simulated ones, with maximum deviations of $\sim$0.5\,Mcps.

These differences highlight the limitations of the analytical model, but also demonstrate its ability to provide reasonable estimates with coarse approximations in the correct order of magnitude. Being aware that the analytical model is just a simple provisional means and not a goal, and that the \MC simulations used as reference might also suffer from large uncertainties \cite{dedes_g4,verburg_simu}, this theoretical evaluation aims at illustrating the fundamental principles and potential feasibility of a new range verification strategy. It does not provide a precise validation or proof-of-principle of \CPGM, but rather encourages its future verification with experimental measurements in a proton therapy facility, and the development of a \MC-based (instead of analytical) detection model, as in \cite{hueso_pgs}.

\subsection{Absolute range prediction}
Our analysis only investigates the sensitivity to range deviations with respect to a nominal well-known situation. A more difficult requirement would be the prediction of the absolute proton range for the first treatment fraction. In that case, the absolute calibration of the ionization chamber measuring the number of incoming protons $N_\mathrm{p}$ is very important \cite{toscano_logs}. Furthermore, the absolute accuracy of the prompt \gr production cross sections would be a critical factor (unless a second detector is added in an upstream location, which can partially rule out a systematic error). Currently, there are deviations between measured and reference cross sections of $\sim$20\% \cite{hueso_pgs}. Hence, an absolute range verification with \CPGM is not realistic in the short to medium term. Extra efforts should be undertaken by a task group within the scientific community, in order to improve the precision down to 2\,\% and to agree on double-differential nuclear reaction cross sections \cite{kiener_angular} in consensus \cite{koegler_nd}.

Furthermore, as the cross sections are target-specific, a precise estimate of the nuclear composition of the irradiated tissue is needed, or it has to be obtained from the ratios between different \gr lines \cite{verburg_spec} and an underlying \MC simulation framework \cite{hueso_pgs}. In this context, it would be also advisable to validate \PGI devices within a standardized experiment framework and ground-truth anthropomorphic phantoms \cite{wohlfahrt_burns}.

Finally, for the particular case of \CPGM, the knowledge about the patient composition behind the irradiated area is also important, as the attenuation of \grs until reaching the coaxial detector also affects the number of \grs measured. This problem will be less critical when dual-energy \CT \cite{wohlfahrt_dect} becomes more commonplace in proton therapy facilities \cite{wohlfahrt_lung}, thereby increasing the precision in the characterization of \gr attenuation coefficients \cite{wohlfahrt_pseudo} and elemental concentrations.

\subsection{Detector count rate}
The choice of LaBr\tlow{3} as scintillation material is motivated by its excellent energy resolution and short decay time of 16\,ns. The first trait is convenient for the compensation of gain fluctations in the \PMT as a function of count rate \cite{werner_pgt}. The second is extremely important in order to minimize the pile-up probability. The expected count rate (including background) for the proposed setup reaches several Mcps, \cf \refig{fig:gammaRate}. This detector load is quite challenging \cite{pausch_needs}, but as only one detector with individual readout is deployed, it can be addressed with sophisticated pile-up fitting algorithms, as shown by \cite{labr_rate,pileup_medical}, and state-of-the-art single-channel digitizers.

\subsection{Radiation background}
The presence of neutron-induced background is a critical factor that might affect the precision of the method. A proportional ratio $k_\gamma$ has been adopted in this study, but experimental measurements are needed to develop a more realistic model depending on the incident proton energy, as well as to verify the absolute yield. Additionally, the increase in neutron fluence in forward direction \cite{zarifi_angle} has to be incorporated in the model. The downstream position of the \CPGM detector is a double-edged sword: it maximizes the sensitivity to prompt \grs near the Bragg peak, but at the price of an increase in neutron-induced background.

A large radiation background is inconvenient for the \CPGM technique due to two reasons. First, it leads to a very high detector count rate, especially at high initial proton energies, that challenges the electronics design. Second, it might bias the estimate of the integral number of prompt \grs, as the subtraction of neutron-induced background based on timing measurements \cite{biegun_tof} is accomplished experimentally with semi-empirical algorithms \cite{hueso_pgs}. Ideally, the neutron-induced background should be fully modeled and incorporated as a measurement value \cite{hauge_neutrons} in addition to prompt \grs, as it carries residual information about the proton range, \cf \refig{fig:gammaDet}. However, the neutron interaction processes are complicated to model and strongly depend on the surrounding materials in the treatment room, which are not well characterized.

\subsection{Integration in treatment room}

\begin{figure}
\centering
\begin{tikzonimage}[width=\columnwidth]{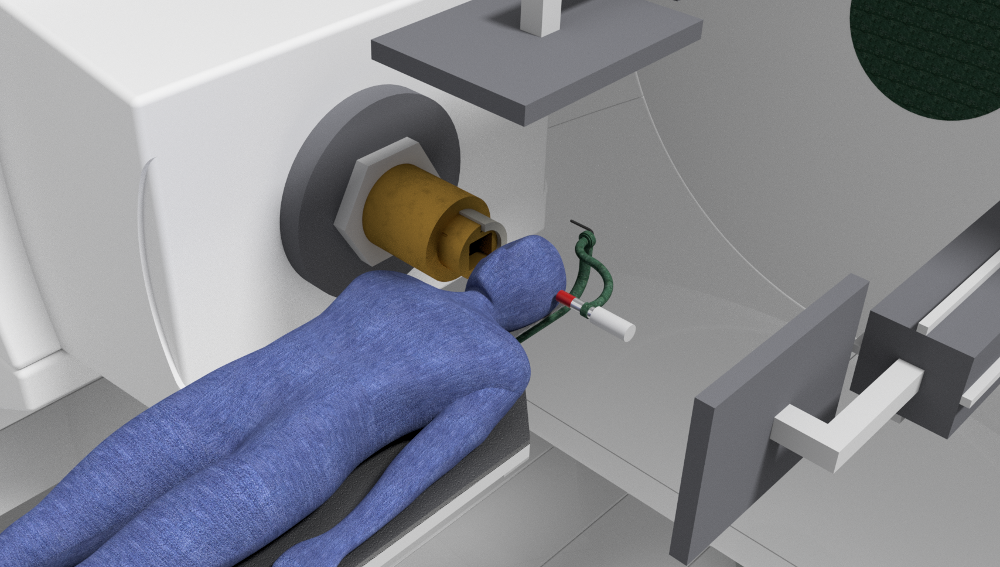}
\sffamily
\small
    \node[align=left, text=red, fill=white,rounded corners, fill opacity=.8, text opacity=1] at (0.24, 0.8) {Nozzle};
    \node[align=left, text=red, fill=white,rounded corners, fill opacity=.8, text opacity=1] at (0.615, 0.35) {Detector};
    \node[align=left, text=red, fill=white,rounded corners, fill opacity=.8, text opacity=1] at (0.38, 0.38) {Patient};
    \node[align=left, text=red, fill=white,rounded corners, fill opacity=.8, text opacity=1] at (0.5, 0.07) {Couch};
    \node[align=left, text=red, fill=white,rounded corners, fill opacity=.8, text opacity=1] at (0.65, 0.62) {Arm};
    \node[align=left, text=red, fill=white,rounded corners, fill opacity=.8, text opacity=1] at (0.8, 0.1) {X-ray panel};
    \node[align=left, text=red, fill=white,rounded corners, fill opacity=.4, text opacity=0.5] at (0.55, 0.90) {X-ray panel};
\end{tikzonimage}
\caption{\ac{3D} model of the proposed detection setup in a proton treatment room. The beam incidence angle is horizontal and the irradiated area is the patient's intracranial region. The scintillation detector (red) is behind the treated area and coaxial to the beam, and is mounted on a rotating arm (green) that is attached to the patient couch. The orthogonal X-ray panels (top and right) can be used for measuring the detector position relative to the patient without interfering the treatment. The number of \gr detections per unit proton is correlated to the beam range. Note that the chosen geometry is compatible with the frequent brain treatment scenario with no couch rotation \cite{susu_gantry}.}
\label{fig:troom}
\end{figure}
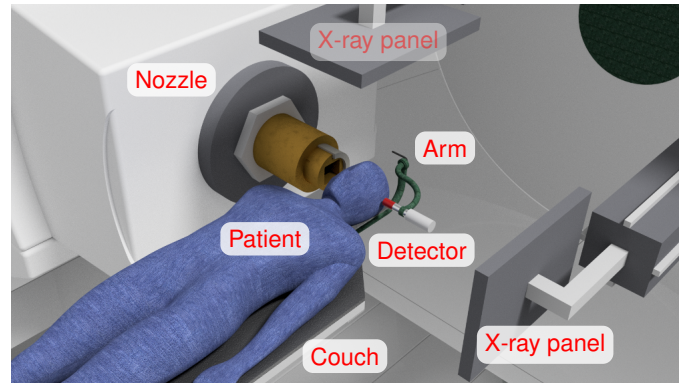

In general, range verification techniques comprise several detection units \cite{werner_pgt_steps,draeger_polaris,krimmer_pgpi,buitenhuis_ibpet}. Some prototypes deploy a thick collimator and a heavy supporting frame \cite[Fig.~10]{hueso_pgs}, \cite[Fig.~4]{petzoldt_geom}, which poses an engineering challenge in case of an integration with the rotating gantry to match any beam angle.
In contrast, the \CPGM is a minimalistic approach using a single detection unit and no collimator.
A model of a gantry treatment room with the \CPGM detector setup attached to the patient couch is illustrated in \refig{fig:troom}. The estimated detector weight is below 500\,g, and the weight of an articulating arm is below 9\,kg. Due to the compactness of the proposed system, its integration seems less complex and more affordable than for other \PGI methods, and is especially promising for treatment rooms with space constraints \cite{slopsema_size}.

\subsection{System alignment}
One important aspect of range verification devices based on \PGI is the system alignment with respect to the room isocenter, \eg with embedded lasers \cite{hueso_pgs}. The positioning uncertainty along the beam direction and the alignment reproducibility are critical, and can be the main limiting factor for the achievable range precision \cite{xie_slit}. Moreover, the heavier the system, the more efforts are needed to achieve the required precision, even more if the \gr camera rotates to match the beam incidence angle.

For the \CPGM approach, we can benefit from the small footprint of the detector as well as its position close to the patient. The in-room orthogonal radiographies used for patient positioning in the clinical workflow are very convenient for detector positioning, as the setup does not collide with the two extended X-ray panels, \cf \refig{fig:troom}. It is also not mandatory to place the detector at the exact same position for every treatment fraction. It is rather necessary to know accurately its position with respect to the patient. The two orthogonal X-ray images can offer a sub-millimeter resolution \cite{sharp_dips} of the \ac{3D} detector position, that can be directly mapped to the beam coordinate system.

\subsection{Compatibility and applicability}
Range verification devices based on \PGI are usually designed for a specific type of proton accelerator used at the developing institution. Consequently, the results cannot be directly translated to other types of accelerators, since the expected beam current and beam time structure \cite{petzoldt_pbm,pausch_needs,walle_s2c2_oma} might be drastically different. In the case of \CPGM, we focus on the most widespread accelerator type \cite{ptcog_facilities}, and the conclusions drawn in this study might potentially not be extrapolated to facilities with a different machine type.

Concerning the versatility of \PGI devices, due to their footprint, a single design cannot be applied in general for every treatment site and every possible patient couch angle so far. Thus, it is convenient to focus first on the most common treatment sites, as well as those who would benefit most of an in vivo range verification. In this study, we choose the treatment of brain tumors from a horizontal incidence angle, and no patient couch rotation, as it is common in clinical practice \cite{susu_gantry}. It should be noted that the study is extensible to other beam incidence angles, as the detector can rotate accordingly around the head, \cf \refig{fig:troom}.

In general terms, the precision of the method increases the closer the detector is to the Bragg peak. Thus, women could potentially benefit more than men from the \CPGM technique due to their smaller average head breadth \cite{zhuang_head_width}. This would be also the case for pediatric tumors.
The extensibility of this approach to other treatment sites depends on the particular beam incidence and patient couch angle. In the case of a prostate treatment from a horizontal beam angle, the coaxial detector position downstream would be far away from the Bragg peak. The range measurement could be done with ease, but the achievable precision would diminish. One alternative could be to move the detector slightly off the beam axis, so that it can reach a closer position to the Bragg peak from above the patient.

\subsection{Price estimate}
The literature about the cost of \PGI prototypes is scarce due to early stage of research development and the several years remaining until a final product is actually commercialized. Some rough estimates state an overall material cost of at least 200\,k\$ \cite{draeger_polaris} and up to 1\,M\$ \cite{rohling_simu_cc}, not accounting for development, commissioning and maintenance. Furthermore, these costs do not include the potential integration in the rotating gantry structure, for covering several beam incidence angles.

The cost-effectiveness of range verification devices is important taking into account its potential applicability in every treatment room of every proton center. Being an optional quality assurance device, not essential for the treatment, the benefits provided by improving the delivery precision should not be burdened by a prohibitive cost in addition to the already expensive proton facility \cite{goitein_costs,peeters_costs}. While there is consensus that \PGI devices are able to provide useful information for proton range determination, it is still questionable whether they will manage to offer it at an affordable price, especially in the case of Compton cameras \cite{hueso_bbcc}. There is a trend towards simpler range verification solutions \cite{golnik_pgt,krimmer_pgpi,parodi_iono} that do not ensure an accuracy as high as already existing prototypes \cite{nenoff_slit,hueso_pgs}, but that might have an easier translation to all clinics due to their lower cost \cite{hueso_bgo}.

In the case of the \CPGM technique, designed for minimum expense and impact on the treatment room, while keeping an acceptable range precision, we provide a budget estimate in table~\ref{tab:budget} based on quotations from year 2019. The total material cost is expected to be about 25\,k\$.

\begin{table}
\begin{tabular}{lr}
    \toprule
    Item                                                & Amount / \$ \\ \midrule

    \textsc{Detector}                                   &              \\
    \phantom{ZZ}\diameter1''$\times$1\nicefrac{1}{2}'' scintillation crystal  &  6,000       \\
    \phantom{ZZ}\diameter1'' fast \PMT                  &  1,000       \\
    \phantom{ZZ}                                        & (7,000)      \\

    \textsc{Electronics}                                &              \\
    \phantom{ZZ}Voltage divider and \HV supply          &  1,000       \\
    \phantom{ZZ}Fast amplifier                          &  1,000       \\
    \phantom{ZZ}Power supply and circuit board          &  1,000       \\
    \phantom{ZZ}                                        & (3,000)      \\

    \textsc{Data acquisition}                           &              \\
    \phantom{ZZ}Fast single-channel digitizer           &  5,000       \\
    \phantom{ZZ}High-end computer                       &  7,000       \\
    \phantom{ZZ}                                        &(12,000)      \\

    \textsc{Mechanics}                                  &              \\
    \phantom{ZZ}Articulating arm                        &  2,000       \\
    \phantom{ZZ}Detector mount                          &  1,000       \\
    \phantom{ZZ}                                        & (3,000)      \\\bottomrule

    \textsc{Total Material Cost}                        & 25,000       \\
\end{tabular}
\caption{Rough estimate of the material costs (budgetary price as of 2019) of a \CPGM device for proton range verification in a clinical treatment room.}
\label{tab:budget}
\end{table}

\subsection{Overview}
Table~\ref{tab:swot} summarizes the \SWOT of the \CPGM method described throughout the manuscript. The simplicity of the design using a single detector is one big advantage. Its small footprint and low weight allow for a seamless integration with the couch of the treatment room for any incidence angle. If a proof-of-principle experiment validates the \CPGM method, it would become an ideal candidate for in vivo range verification in compact proton treatment rooms. At the same time, it does not interfere with the embedded X-ray panels. They could be used for accurate detector positioning, and no additional laser alignment system is needed. Furthermore, the method does not directly depend on the bunch time structure, one of the challenges faced in the \PGT technique \cite{petzoldt_pbm}. Altogether, this minimalist setup might enable range verification at an affordable price of 25\,k\$.

A disadvantage of the \CPGM method is the dependence on the incident number of protons, which is given by the ionization chamber within the beamline. The chamber has to be calibrated very accurately and compensated for pressure or temperature variations. Likewise, the detector response needs to be precisely characterized, and the scintillation crystal aging has to be monitored. Temperature effects on the light yield and \PMT gain should be corrected with stabilization techniques \cite{stein_stabilization}. Furthermore, a good knowledge of the \gr attenuation coefficients in the patient volume behind the treated area is mandatory. The correct estimation of the nuclear concentration within the irradiated area is also essential. Aside from that, another disadvantage is that the distance across the beam axis between detector and treated volume should be as small as possible. Hence, not all treatment sites or patients may benefit from this technique with the same precision. Women and especially children might be more indicated than adult men (on average) for a precise range verification with \CPGM.

Other difficulty faced by the \CPGM technique is the challenging high count rates in the detector, in part due to the increased neutron-induced background in downstream positions. The radiation background has been roughly estimated but has to be verified experimentally. Also, the \CPGM technique depends on nuclear reaction cross section data, which currently suffer from large uncertainties. Aside from that, a potential future obstacle (for any \PGI device) is the increase of the instantaneous beam current in the new accelerators like the synchrocyclotron S2C2 from IBA \cite{walle_s2c2_oma}, that leads to much higher detector count rates. This would force a challenging research towards the use of faster scintillation materials, \eg BaF\tlow{2}, PbWO\tlow{4}, PrBr\tlow{3} \cite{glodo_prbr3} or plastic, than the one deployed in this study, at the price of a worse energy resolution.

Given the aforementioned disadvantages and potential obstacles, there is room for improvement and possible workarounds. One promising way is the addition of a second detector in an upstream position. Both detectors would be sensitive to complementary regions of the patient, which could potentially mitigate errors due to the uncertainty of the nuclear reaction cross sections and even cancel the dependency on the incoming number of protons. A shift of the downstream detector to slightly off-axis angles might also reduce the neutron-induced background, while still keeping a strong dependency of registered \grs with penetration depth. Furthermore, information obtained by \CPGM could be combined with that derived from \PGS or \PGT techniques, but using the same single detector (and no collimator). These can provide additional information to reduce the error in the assumed nuclear concentrations and absolute proton range. Moreover, there might be the possibility to measure the integral number of annihilation photons during beam pauses, as with in-beam \PET \cite{buitenhuis_ibpet}. This could also provide further insights about absolute proton ranges in patients. Of course, a compromise has to be found between the additional information gained by integrating further methods within \CPGM, and the complexity and price that it might add to the method. A simple and robust method with an acceptable accuracy could potentially have more opportunities to become a widespread application in clinic routine.

\newcommand{\textS}{\normalsize Strengths}
\newcommand{\textW}{\normalsize Weaknesses}
\newcommand{\textO}{\normalsize Opportunities}
\newcommand{\textT}{\normalsize Threats}

\newcommand{\texts}{\makebox[0pt][c]{\parbox[t]{0.5\columnwidth}{\centering
Single detector, no collimator\par
Compactness\par
Lightweight\par
Integration with gantry\par
Alignment with X-ray panels\par
Independent of bunch structure\par
Affordable\\[2mm]
}}}
\newcommand{\textw}{\makebox[0pt][c]{\parbox[t]{0.5\columnwidth}{\centering
Dependence on proton number\par
Efficiency calibration\par
Attenuation coefficients\par
Nuclear concentration\par
Shallow treatment sites\\[2mm]
}}}
\newcommand{\texto}{\makebox[0pt][c]{\parbox[t]{0.5\columnwidth}{\centering
Extension to two detectors\par
Slightly off-axis\par
Dual-energy \CT\par
Integration with \PGT\par
Integration with \PGS\par
Combination with \PET\par
Widespread applicability\\[2mm]
}}}
\newcommand{\textt}{\makebox[0pt][c]{\parbox[t]{0.5\columnwidth}{\centering
Challenging count rate\par
Neutron-induced background\par
Nuclear cross sections\par
New accelerators\\[2mm]
}}}

\begin{table}
\centering
\def\arraystretch{2}
\begin{tabular}{*{2}{C{0.45\columnwidth}}}
\cellcolor{helpful} \textS  & \cellcolor{harmful} \textW \\
\mycolor{S}\back{S} \texts & \mycolor{W}\back{W} \textw \\
\mycolor{O}\back{O} \texto & \mycolor{T}\back{T} \textt\\
\cellcolor{opport} \textO  & \cellcolor{threat} \textT \\
\end{tabular}
\vspace{2mm}
\caption{Analysis of \acf{SWOT} of the \CPGM method.}
\label{tab:swot}
\end{table}
 
\section{Conclusions}\label{sec:conclusions}

A method for range verification in proton therapy based on the acquisition of coaxial prompt \grs is presented, which is referred to as \acf{CPGM}. In contrast to other \PGI prototypes comprising several detection units located perpendicularly or upstream of the beam axis, we propose a single \gr detector without collimation positioned coaxially to the proton beam incidence direction. In this particular geometry, and assuming ideal homogeneous phantoms and well-known detector response, the integral number of prompt \grs measured encodes the proton range.

Based on a simplified theoretical model of prompt \gr emission and detection, the sensitivity of \CPGM to virtual range deviations is estimated. For a single $\diameter$1''$\times$1\nicefrac{1}{2}'' LaBr3 scintillation detector positioned behind a 150\,mm water target, we predict a change of more than 3\,\% in the measured number of \grs for a global range error of 1\,mm. Assuming a spot cluster with 10\tup{8} protons, the measurement precision would be enough to detect such a variation with a 1$\sigma$ significance. This compact method could be easily applied in the clinics and would have a seamless integration within the gantry treatment room.

Our theoretical investigation is the first step towards an experimental proof-of-principle study in a proton therapy facility. This experiment has to discern whether the approximations assumed in the model are legitimate and the proposed method can be realistically applied in a clinical scenario. Especially, the predicted neutron-induced background in forward direction and the ability of the detector to cope with very high count rates have to be confirmed. Furthermore, the prompt \gr yields and detector response have to be characterized and calibrated with a very high precision.

Ultimately, the \CPGM method could enable proton range verification at a more affordable price than other \PGI prototypes with heavy collimators and multiple detectors, while keeping an acceptable accuracy. Specifically, we aim at monitoring the proton range deviations with a precision of $\sigma=\,$1\,mm for a spot cluster of 10\tup{8} protons, at a clinical beam current of 2\,nA, setup costs of 25\,k\$ and a weight below 10\,kg. The eventual achievement of this milestone could facilitate a widespread application of in vivo range verification in proton therapy facilities, in order to improve the quality of patient treatment.

\vfill
\section*{Acknowledgement}
We thank P.~Botas, L.~M.~Fraile, C.~Golnik, B.~Gorissen, A.~Hammi, T.~Kögler, H.~Paganetti, G.~Pausch, J.~Petzoldt, T.~A.~Ruggieri, J.~Shin, S.~Tattenberg, J.~M.~Udias, J.~M.~Verburg, A.~Wagner and P.~Wohlfahrt for scientific advice and discussions. This work was supported in part by the Federal Share of program income earned by Massachusetts General Hospital on C06-CA059267, Proton Therapy Research and Treatment Center.

\end{document}